\documentclass[prl,
 reprint,
superscriptaddress,
nofootinbib,
 amsmath,amssymb,
 aps,
]{revtex4-2}

\usepackage{graphicx}
\usepackage{dcolumn}
\usepackage{bm}
\usepackage{hyperref}
\usepackage{upgreek}
\usepackage{booktabs}
\usepackage{tabularx}
\usepackage{comment}
\usepackage{xcolor}

\usepackage{cancel}
\renewcommand{\selectlanguage}[1]{}

\begin{document}
\newcommand{\Eq}{Eq.~}
\newcommand{\Fig}{Fig.~}

\newcommand{\W}{\mathbf{\Omega}}
\newcommand{\nW}{\mathbf{n}_1}
\newcommand{\nV}{\mathbf{n}_2}
\newcommand{\nT}{\mathbf{n}_3}
\newcommand{\nF}{\mathbf{n}_4}
\newcommand{\pos}{\mathbf{r}}
\newcommand{\att}{_{(t)}}
\newcommand{\at}[1]{_{(#1)}}
\newcommand{\expv}[1]{\left\langle #1 \right\rangle}
\newcommand{\D}{\text{d}}
\newcommand{\Dt}{\text{d}t}

\newcommand{\Vo}{\mathbf{V}_0}
\newcommand{\Vomag}{V_0}
\newcommand{\WoVec}{\mathbf{\Omega}_0}
\newcommand{\Wo}{\Omega_0}
\newcommand{\wvec}{\mathbf{\omega}}
\newcommand{\wmag}{\omega}
\newcommand{\norm}[1]{\left\lVert#1\right\rVert}

\newcommand{\rot}{\mathcal{R}}
\newcommand{\dnoise}{\text{d}\mathbf{\Lambda}}
\newcommand{\dscnoise}{\text{d}\Lambda}
\newcommand{\inbody}{^{\text{body}}}
\newcommand{\inlab}{^{\text{lab}}}

\newcommand{\OUth}{k}
\newcommand{\OUamp}{h}

\newcommand{\Van}{\alpha}

\newcommand{\new}[1]{{\leavevmode\color{black}#1}}

\title{Three-dimensional chiral active Ornstein-Uhlenbeck model\\
for helical motion of microorganisms}

\author{Leon Lettermann}
\author{Falko Ziebert}
\affiliation{Institute for Theoretical Physics, Heidelberg University, Philosophenweg 19, 69120 Heidelberg, Germany}
\affiliation{Bioquant-Center, Heidelberg University, Im Neuenheimer Feld 267, 69120 Heidelberg, Germany}
\author{Mirko Singer}
\affiliation{Parasitology, Center for Infectious Diseases, Heidelberg University, Im Neuenheimer Feld 344, 69120 Heidelberg, Germany}
\author{Friedrich Frischknecht}
\affiliation{Parasitology, Center for Infectious Diseases, Heidelberg University, Im Neuenheimer Feld 344, 69120 Heidelberg, Germany}
\affiliation{German Center for Infection Research (DZIF), Partner Site Heidelberg, 69120 Heidelberg, Germany}
\author{Ulrich S. Schwarz}
 \email{schwarz@thphys.uni-heidelberg.de}
\affiliation{Institute for Theoretical Physics, Heidelberg University, Philosophenweg 19, 69120 Heidelberg, Germany}
\affiliation{Bioquant-Center, Heidelberg University, Im Neuenheimer Feld 267, 69120 Heidelberg, Germany}

\date{\today}

\begin{abstract}
Active movement is essential for the survival of microorganisms like
bacteria, algae and unicellular parasites. In three dimensions, both swimming and gliding 
microorganisms often exhibit helical trajectories. 
One such case are malaria parasites gliding through 3D hydrogels, 
for which we find that \new{the internal correlation time for the stochastic process generating propulsion}
is similar to the time taken for one helical turn.
Motivated by this experimental finding, here we theoretically analyze the case of finite internal
correlation time for microorganisms with helical trajectories as 
chiral active particles with an 
Ornstein-Uhlenbeck process for torque.
We present an analytical solution
which is in very good agreement with computer simulations.
We then show that for this type of internal noise, chirality and rotation increase the persistence
of motion and results in helical trajectories that have a larger long-time mean squared displacement
than straight trajectories at the same propulsion speed.
Finally we provide experimental evidence for
this prediction for the case of the malaria parasites.
\end{abstract}

\maketitle

\begin{figure}[b]
    \centering
    \includegraphics[width=\columnwidth]{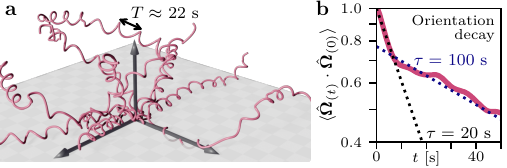}
    \caption{\textbf{a}: Reconstructed trajectories of malaria parasites 
    gliding through synthetic hydrogels. Because this
    environment is nearly isotropic, the right-handed 
    helical trajectories persist for long times. The
    typical (rescaled) turning time $T$ is 22~s as indicated.
    \textbf{b:} The direction of the angular velocity $\hat{\mathbf{\W}}$ displays a 
    decay of autocorrelation with a fast ($\tau=$~20~s) and a slow ($\tau=$~100~s) regime
    \new{(see \Fig\ref{fig:A2} for a version with more details)}.}
    \label{fig:Intro}
\end{figure}

The survival of \new{microorganisms} 
like bacteria or algae is tightly connected to their ability  to actively move,
which is essential to seek out more favorable conditions, e.g. places which offer more nutrients or sunlight for photosynthesis \cite{mitchell2006bacterial, keegstra2022ecological}.
Although sometimes movement is collective, e.g. in biofilms or during swarming,
at the heart of all migration processes is always the capability of
single \new{microorganisms} to internally generate forces and torques \cite{jarrell2008surprisingly, schwarz2015physical, kearns2010field}. Migration over large distances is also essential 
for unicellular eukaryotes that have specialized
in infecting mammalian hosts, such as the causative agents of the diseases
malaria or toxoplasmosis \cite{douglas_cytoskeleton_2024}. Interestingly, in three dimensions many
microorganisms exhibit helical trajectories \cite{jennings_significance_1901,crenshaw_new_1996}.
This includes species of swimming bacteria \cite{berg_chemotaxis_1990, thar_true_2001},
swimming algae \cite{fenchel_motile_1999, jekely_mechanism_2008,leptos2023phototaxis} and 
gliding parasites \cite{ripp_malaria_2021,liu_plasmodium_2024,lettermann_geometrical_2024}. In \Fig\ref{fig:Intro}a,
we show the helical trajectories of malaria parasites gliding through 3D synthetic hydrogels (see Appendix A for details). 

Since the pioneering work of Berg and Purcell, it is well accepted that
one of the main challenges for moving microorganisms is to 
counter the effects of stochastic noise \cite{berg1977physics, berg1993random}. 
For example, the bacterium \textit{E. coli}
uses a run-and-tumble strategy to move up and down
chemotactic gradients; run times are typically of the
order of one second, because for longer times, orientation is lost
due to rotational diffusion \new{which results from collisions with solvent molecules} \cite{pohl2017inferring}. 
The interplay of self-propulsion
and \new{such} external noise can be analyzed by the theory of active
Brownian particles, which combines a \new{constant} propulsion force
with stochastic noise for translation and rotation \cite{romanczuk_active_2012, bechinger_active_2016, zottl_emergent_2016, liebchen2022chiral}. The active Brownian particle model 
has been used and extended in various contexts \cite{howse_self-motile_2007,debnath_diffusion_2016, gomez-solano_dynamics_2016, sevilla2016diffusion, gomez-solano_active_2020, sprenger_dynamics_2023}, 
including adding torques to obtain circle swimmers \cite{van_teeffelen_dynamics_2008, ledesma-aguilar_circle_2012, kummel_circular_2013, marine_diffusive_2013, lowen_chirality_2016, nourhani2016spiral, caprini_chiral_2023} 
and studying the influence of time-correlated noise \cite{lindner_diffusion_2010, weber_active_2011, ghosh_communication_2015, narinder_memory-induced_2018, bayati_memory_2022, sprenger_time-dependent_2021}. 

However, most of these studies considered 2D cases, 
whereas helical motion of chiral active
particles occurs in 3D \cite{sevilla2016diffusion}.
If chiral active motion in 3D was analyzed theoretically,
then mostly in the context of swimming.
An early work on asymmetric swimmers in 3D considered the
analogy to polymer models to extract new power laws for
effective diffusion \cite{rutenberg_diffusion_2003};
in general, the statistics of fluctuating helices
is also an important aspect of helical biopolymers 
like DNA \cite{becker_rigid_2007}.
Previous work on sperm swimming considered 
stochasticity on the level of curvature and torsion
and showed that helical trajectories
are useful search strategies for chemotaxis in noisy
environments \cite{friedrich_stochastic_2008, friedrich_steering_2009}.
This line of work also considered colored noise in the form
of a power spectrum \cite{friedrich_steering_2009}.
One study of chiral motion in 3D started from
the full mobility tensor for an arbitrarily
shaped particle and showed that helical
trajectories are the most likely outcome 
\cite{wittkowski_self-propelled_2012}.
Similarly, a chiral active Brownian particle model has been used to describe 
the helical motion of colonial choanoflagellates and to show that 
purely stochastic propulsion can result in effective dispersion \cite{kirkegaard2016motility}. 
Very recently, it has been shown in a deterministic model
for sperm swimming that an asymmetric beat
of the flagellum leads to helical
trajectories with high persistence
\cite{ren2025swimming}. 
Collectively, this body of work demonstrates
that helical trajectories can have
evolutionary advantages for microorganisms.

Noise does not only arise from the interaction of the
microorganisms with their \new{thermal} environment, but also from the 
internal force generating processes, which might
have a correlation time on the same scale as the
movement that they generate \cite{weber_active_2011}. To address this aspect
of the system, Ornstein-Uhlenbeck (OU) processes have been used,
usually replacing the body-fixed constant velocity with a noisy 
velocity performing an OU-process around a body-fixed average \cite{szamel_self-propelled_2014}.
\new{An OU-process is Brownian motion 
confined by a harmonic potential, such
that the potential minimum defines the average.
The decay time to reach this average 
defines a time scale, thus it effectively corresponds
to colored noise. In recent years, the OU-process 
has been extensively used to model 
the stochastic motion of 2D swimmers}
\cite{sevilla_generalized_2019, woillez_nonlocal_2020, martin_statistical_2021, martin_aoup_2021, nguyen_active_2021, crisanti_most_2023, dutta_most_2023, fritz_thermodynamically_2023}.
However, the approach of using an OU-process
has not yet been explicitly extended to chiral particles in 3D, 
although \new{earlier work addressed a general case of time-correlated
noise with arbitrary power spectrum \cite{friedrich_steering_2009}.
This approach formally includes OU-processes as special case,
but this had not been worked out explicitly before (see below).}

Here we introduce a three-dimensional 
model which represents the noise in the 
generation of torque as an OU-process,
similarly as suggested previously 
for 2D \cite{weber_active_2011}.
This introduces a finite correlation time, reflecting transient but slower additional processes in the torque-generation mechanism, in contrast to uncorrelated (white) Brownian noise.
Analyzing the gliding of the malaria parasites through hydrogels as visualized
in \Fig\ref{fig:Intro}a (see Appendix A for details), we found
that \new{the typical turning time is 22~s and that} the correlation of the direction of the angular velocity
decays with two clearly separated time scales, compare \Fig\ref{fig:Intro}b.
\new{The large time scale of 100~s is much larger than the turning time
and thus corresponds to the decay of orientation 
of the centerline of the helix, which is the global feature
of a helical motion. By contrast, the small time scale of 20~s
must then correspond to more local features in time, in particular
to the correlation time of the force-generating processes.}
The internal correlation time likely results
from reorientation in the flow field of the adhesins on 
the parasite surface, \new{which earlier has been shown both 
experimentally and theoretically to be very variable and
to self-organize for productive gliding 
\cite{hueschen_emergent_2024,lettermann_geometrical_2024}}.

As we demonstrate here, our 3D OU-model for
chiral active particles can be treated
analytically by suitably truncating a
hierarchy of equations. 
We derive equations for the effective correlation time,
the mean position and the mean squared displacement (MSD).
Our main finding is that, in 3D, chirality and hence rotation can lead to enhanced effective persistence compared to non-rotating particles by an integrative effect of stochastic noise; a stabilization that can even lead to helical trajectories becoming "straighter than a straight line", i.e. allowing for larger long-time MSD compared to a particle moving with the same speed on a straight trajectory without rotation.
A similar conclusion has been drawn before from computer simulations of swimming sperm \cite{ren2025swimming}.
This suggests that helical trajectories 
are favored for microorganisms that have to quickly
move large distances through their environment.
Finally we compare our model to experimental data from malaria parasites, 
demonstrating that it can describe the experimentally observed large MSD.

\textit{Model.} \new{The overdamped motion of swimming and gliding microorganisms can be
effectively characterized by translational and angular velocities.
This is equivalent to a description in terms of active forces and
torques for gliding microorganisms \cite{lettermann_geometrical_2024} and has been used before
also for asymmetric low Reynolds number microswimmers \cite{kummel_circular_2013}.}
To model the intrinsic rotational noise, 
we consider an active particle that is moving with 
a body-fixed constant translational velocity 
$\Vo\inbody$. Its rotational velocity performs 
an OU-process around the body-fixed average, 
$\WoVec\inbody$. 
In the lab frame we use two vectors to track 
the orientation of the particle.
$\nW$ is the direction of the mean angular velocity $\WoVec$
and $\nV\perp\nW$ is defined with the 
angle $\alpha$ between $\WoVec$ and $\Vo$ (see
\Fig\ref{fig:Schematics} in the Appendix):
\begin{align}
    \WoVec = \Wo\nW\ , && 
    \Vo= |\Vo|\left(\nW\cos{\Van}+\nV\sin{\Van}\right)\, .
\end{align}
For simplicity, we set $|\Vo|=1$ in the following. 
In the lab frame, the equations of motion are 
\begin{align}
    \D\W\hspace{3pt} =&\ -\OUth \left(\W-\Wo\nW\right)\Dt + \OUamp\dnoise   \label{eq:dOm}\\
    \D\nW =&\ (\W\times\nW) \Dt    \label{eq:dN1}\\
    \D\nV =&\ (\W\times\nV) \Dt    \label{eq:dN2}\\
    \D\pos \hspace{6 pt} =&\ \left(\cos{(\Van)}\nW+\sin{(\Van)}\nV\right)\Dt\,. \label{eq:dr}
\end{align}
Here, $\OUth$ is the potential strength and $\OUamp$ the
noise amplitude of the OU-process. $\dnoise$ is
a 3D standard Wiener process. Note that noise
is not multiplicative in the lab frame. 
Focusing on the intrinsic noise for simplicity, we disregard external noise (such as Brownian translational noise) or intrinsic noise in the translational velocity.

\textit{Rotation.} 
The rotational part described by \Eq\ref{eq:dOm}-\ref{eq:dN1}
is decoupled from the rest and can be solved first.
The dynamical equations for 
the expectation values $\expv{\W}$ and $\expv{\nW}$
constitute an infinite hierarchy of expectation 
values of cross products of these two quantities, 
the first four being $\expv{\W}$, $\expv{\nW}$, $\expv{\W\times\nW}$ and $\expv{\W\times(\W\times\nW)}$.
We can apply moment closure to the higher order terms in the dynamic equation for $\expv{\W\times(\W\times\nW)}$ to truncate this hierarchy (see Appendix B).

By rotational symmetry, only the component singled out by the initially parallel $\W$ and $\nW$ axes is relevant (as the rotational problem is independent of $\nV$), and the other two components of each vector vanish upon averaging (we choose this direction to be $z$). 
Hence, the truncated system defines a four-dimensional, linear, homogeneous ordinary differential equation problem, which we can analyze by its eigenvalues. 

The mode relevant for the long time behavior can be identified as the unique mode with real eigenvalue and parallel $\expv{\W}$ and $\expv{\nW}$, which describes the decorrelation of $\expv{\W}$ from its initial orientation. The other eigenvalues describe the unstable state where $\W$, $\nW$ are anti-parallel, 
and oscillatory states, all of which decay more quickly. The relevant eigenvalue can be
computed exactly, but is cumbersome as a
solution of a fourth-order polynomial.
Expanding for small $\OUth/\Wo$, i.e.~assuming the rotation is faster then the timescale on which the OUP returns to its average, we get the approximation
\begin{equation}\label{eq:eigval}
    \lambda = -\frac{\OUamp^2+\Wo^2 \OUth+\OUth^3-\sqrt{\OUth^2 \left(\Wo^2+\OUth^2\right)^2-\OUamp^4}}{\Wo^2+\OUth^2}\ \,
\end{equation}
which describes the decay as $\expv{\W} = (0,0,\Wo)\exp{(\lambda t)}$ and $\expv{\nW}=(0,0,1)\exp{(\lambda t)}$.
The more negative $\lambda$, the faster $\W$ and $\nW$ decorrelate from their initial orientation. For fixed noise amplitude $h$, both decreasing 
the strength $k$ of the OUP potential and 
decreasing angular speed $\Wo$ lead to faster decorrelation, 
suggesting that the rotation has a stabilizing effect.
In the limits of vanishing noise, 
diverging potential strength or diverging
angular speed,
the time scale of decorrelation diverges.
In the limit of small noise amplitude, $\lambda$ converges to the power spectrum of the OU-process evaluated at the angular speed, consistent with a derivation starting from power spectra (\cite{friedrich_steering_2009}, see Supplemental Note 1).

\begin{figure}
    \centering
    \includegraphics[width=\columnwidth]{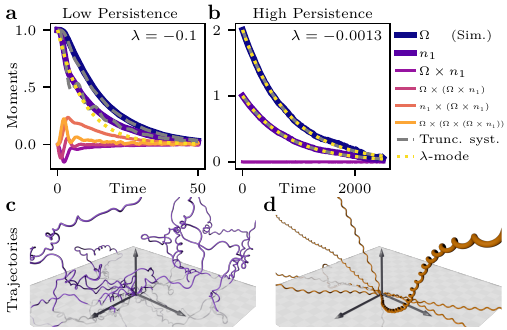}
    \caption{\textbf{a}: Time course of different moments obtained 
    from numerical simulation (\Eq\ref{eq:dOm}-\ref{eq:dr}) in comparison with numerical solution of the truncated system (\Eq\ref{eq:hier1full}-\ref{eq:hier4trunc}, gray dashed lines)
    and analytical approximation predicting exponential
    decay with eigenvalue $\lambda$ (\Eq\ref{eq:eigval}, yellow dotted lines).
    Parameter values: potential strength $\OUth=$~0.2, noise amplitude 
    $\OUamp=$~0.3, angular speed $\Wo=$~1, angle $\Van=\pi/6$.
    \textbf{b}: Same as a, but now for $\OUth=$~2, $\OUamp=$~0.1 and $\Wo=$~2, i.e.~much reduced noise and faster turning. 
    Here, the agreement between simulations and theory is even better.
    \textbf{c}: Simulated trajectories at parameters from a.
    \textbf{d}: Simulated trajectories at parameters from b.
    The reduced noise leads to more regular trajectories. See also Supp. Movies 1+2.
    }
    \label{fig:Rot}
\end{figure}

To validate our approximations, we compared the solutions
against numerical simulations of the initial model (\Eq\ref{eq:dOm}-\ref{eq:dr}, implemented in JAX \cite{bradbury_jax_2018} using standard solvers for stochastic differential equations, set up in diffrax \cite{kidger_neural_2021}). 
In \Fig\ref{fig:Rot}, different expectation values obtained from averaging 20.000 numerical simulations are compared with
(i) the numerically solved truncated ODE system 
(\Eq\ref{eq:hier1full}-\ref{eq:hier4trunc}, dashed gray) and (ii) the
analytical exponential decay given 
by the dominant eigenvalue $\lambda$.
As shown in \Fig\ref{fig:Rot}a, larger noise yielding
faster decorrelation produces larger values of the 
higher order expectation values. We find that the 
truncations are qualitatively correct (see also Supplemental Figure S1), while
quantitative differences are visible --
the numerical solution of the truncated system shows
some additional oscillations.
The exponential decay by $\lambda$ is too fast here, 
which signifies that during the relatively rapid decay,
additional modes are relevant.
For lower noise, in \Fig\ref{fig:Rot}b we see excellent agreement
between numerical simulation, numerical solution of the
truncated system, and the exponential decay given by 
$\lambda$ from \Eq\ref{eq:eigval}. The latter decay completely dictates the
persistence in the resulting motion, as also apparent in the resulting trajectories illustrated in \Fig\ref{fig:Rot}c+d.
Generally, lower decorrelation can be reached by lower noise amplitude $\OUamp$, stronger Ornstein-Uhlenbeck potential $\OUth$, or higher angular speed $\Wo$.\\

\textit{Translation.}
For the analytical treatment of the translational part, we assume that
initially $\W=\Wo\nW$ is in $z$ direction. The solution of 
the rotational part then allows to solve 
\Eq\ref{eq:dr} for the motion in $z$,
$\D\expv{z} =\ \cos{(\Van)} \expv{\nW}$. 
To obtain the MSD and the remaining coordinate, 
we need an expression for $\expv{\nV}$. 
By construction, $\nV\perp\nW$, so $\nV$ is rotating in the plane perpendicular to $\nW$ with angular frequency $\Wo$, which is on average the $x$-$y$-plane.
We assume that $\nV$ initially points in $x$-direction.
The decorrelation of $\nW$ is also decorrelating the plane in which $\nV$ rotates, but the latter additionally decorrelates within the plane by variations of the magnitude of the rotational velocity. Both effects are caused by $\W$ deviating from $\WoVec$, the tilting of the plane by deviations perpendicular to $\WoVec$, and in-plane deviations by parallel components. Because of this additional effect, we obtain a two-fold faster decorrelation of $\expv{\nV}$ compared to $\expv{\nW}$ (see Supplemental Note 2),
\begin{equation}\label{eq:solnV}
    \expv{\nV}=\big(\cos{\Wo t},\sin{\Wo t},0\big)\,e^{2\lambda t}\ .
\end{equation}
The MSD can now be obtained (see Appendix C) by first computing it from the formal solution
and inserting the solutions obtained for $\expv{\nW}$ and $\expv{\nV}$. The result is (except for the degenerate case $\Wo=0$ and $\Van>0$):
\begin{equation}\label{eq:MSD}
    \begin{split}
        &\expv{\pos^2\att}= \frac{2 \cos ^2(\Van) \left(-\lambda t+e^{\lambda t}-1\right)}{\lambda^2}\\
        &\ +\frac{2 \sin ^2(\Van)}{\left(4 \lambda^2+\Wo^2\right)^2} \Big[-4 \lambda^2+\Wo^2-2 \lambda\left(4\lambda^2 +\Wo^2\right) t\\
        &\ +\left(4 \lambda^2-\Wo^2\right) e^{2 \lambda t} \cos (\Wo t)+4 \lambda \Wo e^{2 \lambda t} \sin (\Wo t)\Big]\ .
    \end{split}
\end{equation}
Let us consider two limiting cases.
First, for $\Van=0$, corresponding 
to a particle rotating while traveling in average on a straight line, we obtain
\begin{equation}
\expv{\pos\att^2} = -\frac{2}{\lambda}t-\frac{2}{\lambda^2}\left(1-e^{\lambda t}\right)
\ ,
\end{equation}
which recovers the case of an active Brownian particle (see Supplemental Note 1, note $\lambda<0$).
Second, for general $\alpha$ 
in the limit of large $t$, we can approximate $\expv{\pos\att^2} \approx 6 D_\infty t$,
where we obtain the effective diffusion 
constant describing the long time behavior as
\begin{equation}\label{eq:Diffconst}
    D_\infty = -\frac{\lambda}{3} \left(\frac{\cos ^2(\Van)}{\lambda^2}+\frac{2\sin ^2(\Van)}{\left(4\lambda^2+\Wo^2\right)}\right) \ .
\end{equation}
In the case 
$\lambda^2\ll\Wo^2$, meaning small noise 
leading to a decay time much longer 
than the rotation period, 
and $\Van<\pi/2$, i.e.~the particle not just circling, but having some average net movement, this reduces to $D_\infty \approx -\cos ^2(\Van)/(3 \lambda)$.

Lastly, with \Eq\ref{eq:solnV} we compute $\expv{x}$ and $\expv{y}$ by integrating
    $\D\expv{x} =\sin (\Van) \expv{\nV}_x \Dt $,
allowing to obtain the expectation value of 
the trajectory,
\begin{equation}\label{eq:expvalPos}
    \expv{\pos\att} = \begin{pmatrix}
        \sin (\Van) \frac{e^{2 \lambda t} (2 \lambda \cos (\Wo t)+\Wo \sin (\Wo t))-2 \lambda}{4 \lambda^2+\Wo^2}\\
        \sin (\Van) \frac{e^{2 \lambda t} (2 \lambda \sin (\Wo t)-\Wo \cos (\Wo t))+\Wo}{4 \lambda^2+\Wo^2}\\
        \frac{\cos{(\Van)}}{\lambda}\left(e^{\lambda t} - 1\right)
    \end{pmatrix}\ ,
\end{equation}
which is a logarithmic spiral on a 
radical surface, i.e.~$z\propto\sqrt{r}$.\\

\begin{figure}
    \centering
    \includegraphics[width=\columnwidth]{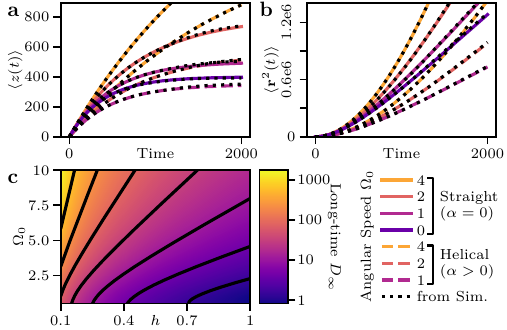}
    \caption{\textbf{a}: Mean distance traveled in 
    $z$-direction (the initial orientation of the helical axis) for different $\Wo$ at $\OUth=$~2, $\OUamp=$~0.1. 
    Full lines show particles moving straight while turning ($\Van$=0), dashed lines particles on helical trajectories ($\Van$=$\pi/4$), which can be seen overtaking slower turning straight particles. Colored and black lines are theoretical and numerical,
    respectively, and in very good agreement. 
    \textbf{b}: Mean squared displacement for the same parameters as shown in a, theoretical results from \Eq\ref{eq:MSD} in color.
    \textbf{c}: Effective long-time diffusion constant $D_\infty$, cf.~\Eq\ref{eq:Diffconst}, as a function 
    of noise amplitude $\OUamp$ and angular speed $\Wo$. Black lines mark contours of constant $D_\infty$.
    }
    \label{fig:Trans}
\end{figure}

\textit{Results.}
The derived solutions show that
increasing rotation $\Wo$ stabilizes 
the particle against its intrinsic noise.
In \Fig\ref{fig:Trans}a and b we plot
the mean $z$-position and the MSD, respectively, 
which both increase with increasing angular speed. 
In both cases we find that the numerical 
simulations agree well with the analytical results,
cf.~the third component of \Eq\ref{eq:expvalPos}
and \Eq\ref{eq:MSD}. The plots show that
if the particle travels on a helix 
(case $\Van=\pi/4$, cf.~dashed lines in 
\Fig\ref{fig:Trans}a,b) 
with the same speed as a non-rotating particle
traveling in a straight fashion, 
if it turns sufficiently fast 
(i.e.~if the helix is sufficiently tightly wound) it travels further from the origin on average at large time scales.
Therefore a helical trajectory can be "straighter than
a straight line".

The long time behavior is described by the effective diffusion 
constant \Eq\ref{eq:Diffconst}, which has a complicated dependence on $\Wo$, $\OUth$ and $\OUamp$ through $\lambda$. 
In \Fig\ref{fig:Trans}c we see that at constant 
OU potential strength $\OUth$, $D_\infty$ increases with higher angular speed $\Wo$, as this suppresses deviations of the helical axis, different from what was found for chiral active Brownian particles 
without the OU-process \cite{kirkegaard2016motility} (see Supplemental Note 1).
This effect becomes more pronounced for higher noise amplitude $\OUamp$, i.e. at higher noise, the stabilizing effect of rotation is more pronounced. 
Increasing effective diffusion by introducing rotation or equivalently chirality is strikingly different from known examples. In 2D, chirality reduces long-time diffusion by enforcing circular turning \cite{lowen_chirality_2016, olsen2024optimal}.
Similarly, a 3D active Brownian particle with external torque exhibits reduced long-time diffusion \cite{sevilla2016diffusion}.

\begin{figure}
    \centering
    \includegraphics[width=\columnwidth]{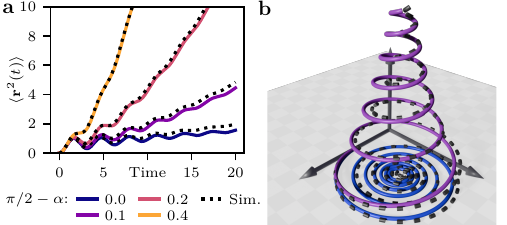}
    \caption{\textbf{a}: Mean squared displacement for $\Van$ close to $\pi/2$, such that the particles are close to describing circles, with $\OUth=$~1, $\OUamp=$~0.5, $\Wo=$~2. Black dotted lines are averages from numerical simulations.
    \textbf{b}: Theoretical expectation value of trajectories (\Eq\ref{eq:expvalPos}) for the two lower values of $\pi/2-\Van$.}
    \label{fig:Fig4}
\end{figure}

\begin{figure}[b]
    \centering
    \includegraphics[width=\columnwidth]{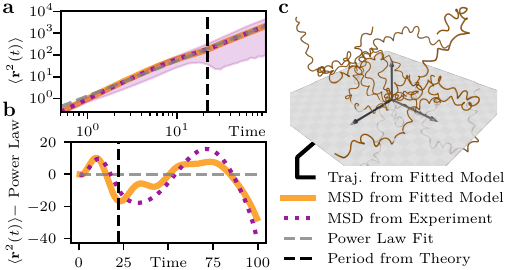}
    \caption{\textbf{a}: Log-log plot of the mean squared displacement extracted from observed malaria parasite trajectories shown in \Fig\ref{fig:Intro} (purple, dotted), with five percent percentiles (purple, shaded) and the fitted model (orange). The gray dashed line is a fitted power law.
    \textbf{b}: Deviation from fitted power law. The vertical dashed line marks one period of rotation as extracted from the fitted model.
    \textbf{c}: Trajectories simulated with parameters obtained from MSD fit
    resembling \Fig\ref{fig:Intro} (cf. Supp. Movie 3).}
    \label{fig:Fig5}
\end{figure}

We can also study the short time behavior.
\Fig\ref{fig:Fig4}a shows that
at short times, the MSD of a helix grows 
slower, because it is curving back onto
itself, depending on the pitch of the
helix defined by the angle $\Van$. 
For $\Van=\pi/2$, the MSD shows strong
oscillations, as the mean position, 
see \Fig\ref{fig:Fig4}b, describes a planar 
inward spiral due to the influence of noise 
that diverts it from the circle of a noise-free
particle. For smaller $\Van$, the spiral gets
the 3D structure of a logarithmic spiral 
on a radical surface as found in
\Eq\ref{eq:expvalPos}, with both cases showing 
good agreement between the numerical and
analytical results (a similar spiral was found numerically
in \cite{wittkowski_self-propelled_2012} for active Brownian particles
with torque, see Supplemental Note 1).

Finally, we can use our measured trajectories for malaria
parasites in hydrogels to extract their MSD (\Fig\ref{fig:Fig5}a, averaged from 140 trajectories) and fit it with our model prediction of \Eq\ref{eq:MSD}, similar to what has been done before in 2D-projections for choanoflagellate colonies \cite{kirkegaard2016motility}. In general, we find good agreement. Our theory successfully describes the first two extrema in the deviation
of the MSD from a power law (\Fig\ref{fig:Fig5}b), corresponding to the first turn of the helix. Our theory also predicts some effects of second and third turns visible in the MSD deviation, which are not observable in the experimental data, most likely because the biological population has a distribution of helical pitches and radii such that the later turns cannot be resolved in the average. From the fitted model parameters (Appendix A), we can derive estimates for pitch and radius of the helical trajectories as 13.2~$\upmu$m and 2.8~$\upmu$m, respectively, well within the observed range \cite{ripp_malaria_2021}. We note in passing
that a phenomenological fit of the period as the period in the MSD-deviation would yield incorrect results for the oscillation period. Trajectories simulated with the fitted parameters (\Fig\ref{fig:Fig5}c) visually resemble the observed trajectories from \Fig\ref{fig:Intro}.

In summary, our results suggest that helical trajectories
provide an advantage for swimming or
gliding microorganisms with noisy force generation
to effectively cover distance more
quickly than 
when going straight. This result 
from a general stochastic theory
for microorganisms with colored noise
in their internal torque generation
complements early insight about the
potential evolutionary advantages
of helical motion for swimming \cite{friedrich_stochastic_2008,kirkegaard2016motility,ren2025swimming}. 
\new{Our results should apply also to cases of microswimmers with large internal correlation
times, like the case of {\it Chlamydomonas} \cite{wan_rhythmicity_2014}. 
In general, it would be highly interesting to
correlate measured correlation times with motion patterns and 
to interpret them in their ecological and evolutionary contexts.}
In the future, our model could also guide the design of
micro- and nanobots \cite{ghosh_controlled_2009, yamamoto2017chirality}, for example in medical applications where enhanced persistence of motion is required \cite{llacer2021biodegradable}.

\textit{Acknowledgments:} This work was funded by the Deutsche Forschungsgemeinschaft (DFG, German Research Foundation) through Priority Programme 2332 (Projektnummer 492010213) and Collaborative Research Center 1129 (Projektnummer 240245660).

\textit{Data availability:} The code used to simulate the model, as well as python implementations of the analytical solutions for comparison are available in the following repository: GITHUB URL UPON PUBLICATION

\bigskip

\bibliography{references}

\begin{thebibliography}{11}%
\makeatletter
\providecommand \@ifxundefined [1]{%
 \@ifx{#1\undefined}
}%
\providecommand \@ifnum [1]{%
 \ifnum #1\expandafter \@firstoftwo
 \else \expandafter \@secondoftwo
 \fi
}%
\providecommand \@ifx [1]{%
 \ifx #1\expandafter \@firstoftwo
 \else \expandafter \@secondoftwo
 \fi
}%
\providecommand \natexlab [1]{#1}%
\providecommand \enquote  [1]{``#1''}%
\providecommand \bibnamefont  [1]{#1}%
\providecommand \bibfnamefont [1]{#1}%
\providecommand \citenamefont [1]{#1}%
\providecommand \href@noop [0]{\@secondoftwo}%
\providecommand \href [0]{\begingroup \@sanitize@url \@href}%
\providecommand \@href[1]{\@@startlink{#1}\@@href}%
\providecommand \@@href[1]{\endgroup#1\@@endlink}%
\providecommand \@sanitize@url [0]{\catcode `\\12\catcode `\$12\catcode `\&12\catcode `\#12\catcode `\^12\catcode `\_12\catcode `\%12\relax}%
\providecommand \@@startlink[1]{}%
\providecommand \@@endlink[0]{}%
\providecommand \url  [0]{\begingroup\@sanitize@url \@url }%
\providecommand \@url [1]{\endgroup\@href {#1}{\urlprefix }}%
\providecommand \urlprefix  [0]{URL }%
\providecommand \Eprint [0]{\href }%
\providecommand \doibase [0]{https://doi.org/}%
\providecommand \selectlanguage [0]{\@gobble}%
\providecommand \bibinfo  [0]{\@secondoftwo}%
\providecommand \bibfield  [0]{\@secondoftwo}%
\providecommand \translation [1]{[#1]}%
\providecommand \BibitemOpen [0]{}%
\providecommand \bibitemStop [0]{}%
\providecommand \bibitemNoStop [0]{.\EOS\space}%
\providecommand \EOS [0]{\spacefactor3000\relax}%
\providecommand \BibitemShut  [1]{\csname bibitem#1\endcsname}%
\let\auto@bib@innerbib\@empty
\bibitem [{\citenamefont {Howse}\ \emph {et~al.}(2007)\citenamefont {Howse}, \citenamefont {Jones}, \citenamefont {Ryan}, \citenamefont {Gough}, \citenamefont {Vafabakhsh},\ and\ \citenamefont {Golestanian}}]{howse_self-motile_2007}%
  \BibitemOpen
  \bibfield  {author} {\bibinfo {author} {\bibfnamefont {J.~R.}\ \bibnamefont {Howse}}, \bibinfo {author} {\bibfnamefont {R.~A.~L.}\ \bibnamefont {Jones}}, \bibinfo {author} {\bibfnamefont {A.~J.}\ \bibnamefont {Ryan}}, \bibinfo {author} {\bibfnamefont {T.}~\bibnamefont {Gough}}, \bibinfo {author} {\bibfnamefont {R.}~\bibnamefont {Vafabakhsh}},\ and\ \bibinfo {author} {\bibfnamefont {R.}~\bibnamefont {Golestanian}},\ }\bibfield  {title} {\bibinfo {title} {Self-{Motile} {Colloidal} {Particles}: {From} {Directed} {Propulsion} to {Random} {Walk}},\ }\href@noop {} {\bibfield  {journal} {\bibinfo  {journal} {Phys. Rev. Lett.}\ }\textbf {\bibinfo {volume} {99}},\ \bibinfo {pages} {048102} (\bibinfo {year} {2007})}\BibitemShut {NoStop}%
\bibitem [{\citenamefont {Friedrich}\ and\ \citenamefont {Jülicher}(2009)}]{friedrich_steering_2009}%
  \BibitemOpen
  \bibfield  {author} {\bibinfo {author} {\bibfnamefont {B.~M.}\ \bibnamefont {Friedrich}}\ and\ \bibinfo {author} {\bibfnamefont {F.}~\bibnamefont {Jülicher}},\ }\bibfield  {title} {\bibinfo {title} {Steering {Chiral} {Swimmers} along {Noisy} {Helical} {Paths}},\ }\href@noop {} {\bibfield  {journal} {\bibinfo  {journal} {Phys. Rev. Lett.}\ }\textbf {\bibinfo {volume} {103}},\ \bibinfo {pages} {068102} (\bibinfo {year} {2009})}\BibitemShut {NoStop}%
\bibitem [{\citenamefont {Kirkegaard}\ \emph {et~al.}(2016)\citenamefont {Kirkegaard}, \citenamefont {Marron},\ and\ \citenamefont {Goldstein}}]{kirkegaard2016motility}%
  \BibitemOpen
  \bibfield  {author} {\bibinfo {author} {\bibfnamefont {J.~B.}\ \bibnamefont {Kirkegaard}}, \bibinfo {author} {\bibfnamefont {A.~O.}\ \bibnamefont {Marron}},\ and\ \bibinfo {author} {\bibfnamefont {R.~E.}\ \bibnamefont {Goldstein}},\ }\bibfield  {title} {\bibinfo {title} {Motility of colonial choanoflagellates and the statistics of aggregate random walkers},\ }\href@noop {} {\bibfield  {journal} {\bibinfo  {journal} {Phys. Rev. Lett.}\ }\textbf {\bibinfo {volume} {116}},\ \bibinfo {pages} {038102} (\bibinfo {year} {2016})}\BibitemShut {NoStop}%
\bibitem [{\citenamefont {Sevilla}(2016)}]{sevilla2016diffusion}%
  \BibitemOpen
  \bibfield  {author} {\bibinfo {author} {\bibfnamefont {F.~J.}\ \bibnamefont {Sevilla}},\ }\bibfield  {title} {\bibinfo {title} {Diffusion of active chiral particles},\ }\href@noop {} {\bibfield  {journal} {\bibinfo  {journal} {Phys. Rev. E}\ }\textbf {\bibinfo {volume} {94}},\ \bibinfo {pages} {062120} (\bibinfo {year} {2016})}\BibitemShut {NoStop}%
\bibitem [{\citenamefont {Kratky}\ and\ \citenamefont {Porod}(1949)}]{kratky1949rontgenuntersuchung}%
  \BibitemOpen
  \bibfield  {author} {\bibinfo {author} {\bibfnamefont {O.}~\bibnamefont {Kratky}}\ and\ \bibinfo {author} {\bibfnamefont {G.}~\bibnamefont {Porod}},\ }\bibfield  {title} {\bibinfo {title} {R{\"o}ntgenuntersuchung gel{\"o}ster fadenmolek{\"u}le},\ }\href@noop {} {\bibfield  {journal} {\bibinfo  {journal} {Recueil des Travaux Chimiques des Pays-Bas}\ }\textbf {\bibinfo {volume} {68}},\ \bibinfo {pages} {1106} (\bibinfo {year} {1949})}\BibitemShut {NoStop}%
\bibitem [{\citenamefont {Doi}\ \emph {et~al.}(1988)\citenamefont {Doi}, \citenamefont {Edwards},\ and\ \citenamefont {Edwards}}]{doi1988theory}%
  \BibitemOpen
  \bibfield  {author} {\bibinfo {author} {\bibfnamefont {M.}~\bibnamefont {Doi}}, \bibinfo {author} {\bibfnamefont {S.~F.}\ \bibnamefont {Edwards}},\ and\ \bibinfo {author} {\bibfnamefont {S.~F.}\ \bibnamefont {Edwards}},\ }\href@noop {} {\emph {\bibinfo {title} {The theory of polymer dynamics}}},\ Vol.~\bibinfo {volume} {73}\ (\bibinfo  {publisher} {Oxford University Press},\ \bibinfo {year} {1988})\BibitemShut {NoStop}%
\bibitem [{\citenamefont {Wittkowski}\ and\ \citenamefont {Löwen}(2012)}]{wittkowski_self-propelled_2012}%
  \BibitemOpen
  \bibfield  {author} {\bibinfo {author} {\bibfnamefont {R.}~\bibnamefont {Wittkowski}}\ and\ \bibinfo {author} {\bibfnamefont {H.}~\bibnamefont {Löwen}},\ }\bibfield  {title} {\bibinfo {title} {Self-propelled {Brownian} spinning top: {Dynamics} of a biaxial swimmer at low {Reynolds} numbers},\ }\href@noop {} {\bibfield  {journal} {\bibinfo  {journal} {Phys. Rev. E}\ }\textbf {\bibinfo {volume} {85}},\ \bibinfo {pages} {021406} (\bibinfo {year} {2012})}\BibitemShut {NoStop}%
\bibitem [{\citenamefont {Friedrich}\ and\ \citenamefont {Jülicher}(2008)}]{friedrich_stochastic_2008}%
  \BibitemOpen
  \bibfield  {author} {\bibinfo {author} {\bibfnamefont {B.~M.}\ \bibnamefont {Friedrich}}\ and\ \bibinfo {author} {\bibfnamefont {F.}~\bibnamefont {Jülicher}},\ }\bibfield  {title} {{\selectlanguage {en}\bibinfo {title} {The stochastic dance of circling sperm cells: sperm chemotaxis in the plane}},\ }\href@noop {} {\bibfield  {journal} {\bibinfo  {journal} {New J. Phys.}\ }\textbf {\bibinfo {volume} {10}},\ \bibinfo {pages} {123025} (\bibinfo {year} {2008})}\BibitemShut {NoStop}%
\bibitem [{\citenamefont {Gardiner}(1985)}]{gardiner1985handbook}%
  \BibitemOpen
  \bibfield  {author} {\bibinfo {author} {\bibfnamefont {C.~W.}\ \bibnamefont {Gardiner}},\ }\bibfield  {title} {\bibinfo {title} {Handbook of stochastic methods for physics, chemistry and the natural sciences},\ }\href@noop {} {\bibfield  {journal} {\bibinfo  {journal} {Springer series in synergetics}\ } (\bibinfo {year} {1985})}\BibitemShut {NoStop}%
\bibitem [{\citenamefont {H{\"a}nggi}\ and\ \citenamefont {Jung}(1994)}]{hanggi1994colored}%
  \BibitemOpen
  \bibfield  {author} {\bibinfo {author} {\bibfnamefont {P.}~\bibnamefont {H{\"a}nggi}}\ and\ \bibinfo {author} {\bibfnamefont {P.}~\bibnamefont {Jung}},\ }\bibfield  {title} {\bibinfo {title} {Colored noise in dynamical systems},\ }\href@noop {} {\bibfield  {journal} {\bibinfo  {journal} {Adv. Chem. Phys.}\ }\textbf {\bibinfo {volume} {89}},\ \bibinfo {pages} {239} (\bibinfo {year} {1994})}\BibitemShut {NoStop}%
\bibitem [{\citenamefont {Bhattacharya}(1982)}]{bhattacharya_functional_1982}%
  \BibitemOpen
  \bibfield  {author} {\bibinfo {author} {\bibfnamefont {R.~N.}\ \bibnamefont {Bhattacharya}},\ }\bibfield  {title} {{\selectlanguage {en}\bibinfo {title} {On the functional central limit theorem and the law of the iterated logarithm for {Markov} processes}},\ }\href@noop {} {\bibfield  {journal} {\bibinfo  {journal} {Z. Wahrscheinlichkeitstheorie Verw. Geb.}\ }\textbf {\bibinfo {volume} {60}},\ \bibinfo {pages} {185} (\bibinfo {year} {1982})}\BibitemShut {NoStop}%
\end{thebibliography}%


\begin{thebibliography}{67}%
\makeatletter
\providecommand \@ifxundefined [1]{%
 \@ifx{#1\undefined}
}%
\providecommand \@ifnum [1]{%
 \ifnum #1\expandafter \@firstoftwo
 \else \expandafter \@secondoftwo
 \fi
}%
\providecommand \@ifx [1]{%
 \ifx #1\expandafter \@firstoftwo
 \else \expandafter \@secondoftwo
 \fi
}%
\providecommand \natexlab [1]{#1}%
\providecommand \enquote  [1]{``#1''}%
\providecommand \bibnamefont  [1]{#1}%
\providecommand \bibfnamefont [1]{#1}%
\providecommand \citenamefont [1]{#1}%
\providecommand \href@noop [0]{\@secondoftwo}%
\providecommand \href [0]{\begingroup \@sanitize@url \@href}%
\providecommand \@href[1]{\@@startlink{#1}\@@href}%
\providecommand \@@href[1]{\endgroup#1\@@endlink}%
\providecommand \@sanitize@url [0]{\catcode `\\12\catcode `\$12\catcode `\&12\catcode `\#12\catcode `\^12\catcode `\_12\catcode `\%12\relax}%
\providecommand \@@startlink[1]{}%
\providecommand \@@endlink[0]{}%
\providecommand \url  [0]{\begingroup\@sanitize@url \@url }%
\providecommand \@url [1]{\endgroup\@href {#1}{\urlprefix }}%
\providecommand \urlprefix  [0]{URL }%
\providecommand \Eprint [0]{\href }%
\providecommand \doibase [0]{https://doi.org/}%
\providecommand \selectlanguage [0]{\@gobble}%
\providecommand \bibinfo  [0]{\@secondoftwo}%
\providecommand \bibfield  [0]{\@secondoftwo}%
\providecommand \translation [1]{[#1]}%
\providecommand \BibitemOpen [0]{}%
\providecommand \bibitemStop [0]{}%
\providecommand \bibitemNoStop [0]{.\EOS\space}%
\providecommand \EOS [0]{\spacefactor3000\relax}%
\providecommand \BibitemShut  [1]{\csname bibitem#1\endcsname}%
\let\auto@bib@innerbib\@empty
\bibitem [{\citenamefont {Mitchell}\ and\ \citenamefont {Kogure}(2006)}]{mitchell2006bacterial}%
  \BibitemOpen
  \bibfield  {author} {\bibinfo {author} {\bibfnamefont {J.~G.}\ \bibnamefont {Mitchell}}\ and\ \bibinfo {author} {\bibfnamefont {K.}~\bibnamefont {Kogure}},\ }\bibfield  {title} {\bibinfo {title} {Bacterial motility: links to the environment and a driving force for microbial physics},\ }\href@noop {} {\bibfield  {journal} {\bibinfo  {journal} {FEMS Microbiol. Ecol.}\ }\textbf {\bibinfo {volume} {55}},\ \bibinfo {pages} {3} (\bibinfo {year} {2006})}\BibitemShut {NoStop}%
\bibitem [{\citenamefont {Keegstra}\ \emph {et~al.}(2022)\citenamefont {Keegstra}, \citenamefont {Carrara},\ and\ \citenamefont {Stocker}}]{keegstra2022ecological}%
  \BibitemOpen
  \bibfield  {author} {\bibinfo {author} {\bibfnamefont {J.~M.}\ \bibnamefont {Keegstra}}, \bibinfo {author} {\bibfnamefont {F.}~\bibnamefont {Carrara}},\ and\ \bibinfo {author} {\bibfnamefont {R.}~\bibnamefont {Stocker}},\ }\bibfield  {title} {\bibinfo {title} {The ecological roles of bacterial chemotaxis},\ }\href@noop {} {\bibfield  {journal} {\bibinfo  {journal} {Nat. Rev. Microbiol}\ }\textbf {\bibinfo {volume} {20}},\ \bibinfo {pages} {491} (\bibinfo {year} {2022})}\BibitemShut {NoStop}%
\bibitem [{\citenamefont {Jarrell}\ and\ \citenamefont {McBride}(2008)}]{jarrell2008surprisingly}%
  \BibitemOpen
  \bibfield  {author} {\bibinfo {author} {\bibfnamefont {K.~F.}\ \bibnamefont {Jarrell}}\ and\ \bibinfo {author} {\bibfnamefont {M.~J.}\ \bibnamefont {McBride}},\ }\bibfield  {title} {\bibinfo {title} {The surprisingly diverse ways that prokaryotes move},\ }\href@noop {} {\bibfield  {journal} {\bibinfo  {journal} {Nat. Rev. Microbiol}\ }\textbf {\bibinfo {volume} {6}},\ \bibinfo {pages} {466} (\bibinfo {year} {2008})}\BibitemShut {NoStop}%
\bibitem [{\citenamefont {Schwarz}(2015)}]{schwarz2015physical}%
  \BibitemOpen
  \bibfield  {author} {\bibinfo {author} {\bibfnamefont {U.~S.}\ \bibnamefont {Schwarz}},\ }\bibfield  {title} {\bibinfo {title} {Physical constraints for pathogen movement},\ }in\ \href@noop {} {\emph {\bibinfo {booktitle} {Semin. Cell Dev. Biol.}}},\ Vol.~\bibinfo {volume} {46}\ (\bibinfo {organization} {Elsevier},\ \bibinfo {year} {2015})\ pp.\ \bibinfo {pages} {82--90}\BibitemShut {NoStop}%
\bibitem [{\citenamefont {Kearns}(2010)}]{kearns2010field}%
  \BibitemOpen
  \bibfield  {author} {\bibinfo {author} {\bibfnamefont {D.~B.}\ \bibnamefont {Kearns}},\ }\bibfield  {title} {\bibinfo {title} {A field guide to bacterial swarming motility},\ }\href@noop {} {\bibfield  {journal} {\bibinfo  {journal} {Nat. Rev. Microbiol}\ }\textbf {\bibinfo {volume} {8}},\ \bibinfo {pages} {634} (\bibinfo {year} {2010})}\BibitemShut {NoStop}%
\bibitem [{\citenamefont {Douglas}\ \emph {et~al.}(2024)\citenamefont {Douglas}, \citenamefont {Moon},\ and\ \citenamefont {Frischknecht}}]{douglas_cytoskeleton_2024}%
  \BibitemOpen
  \bibfield  {author} {\bibinfo {author} {\bibfnamefont {R.~G.}\ \bibnamefont {Douglas}}, \bibinfo {author} {\bibfnamefont {R.~W.}\ \bibnamefont {Moon}},\ and\ \bibinfo {author} {\bibfnamefont {F.}~\bibnamefont {Frischknecht}},\ }\bibfield  {title} {{\selectlanguage {eng}\bibinfo {title} {Cytoskeleton {Organization} in {Formation} and {Motility} of {Apicomplexan} {Parasites}}},\ }\href@noop {} {\bibfield  {journal} {\bibinfo  {journal} {Annu. Rev. Microbiol}\ }\textbf {\bibinfo {volume} {78}},\ \bibinfo {pages} {311} (\bibinfo {year} {2024})}\BibitemShut {NoStop}%
\bibitem [{\citenamefont {Jennings}(1901)}]{jennings_significance_1901}%
  \BibitemOpen
  \bibfield  {author} {\bibinfo {author} {\bibfnamefont {H.~S.}\ \bibnamefont {Jennings}},\ }\bibfield  {title} {\bibinfo {title} {On the {Significance} of the {Spiral} {Swimming} of {Organisms}},\ }\href@noop {} {\bibfield  {journal} {\bibinfo  {journal} {Am. Nat.}\ }\textbf {\bibinfo {volume} {35}},\ \bibinfo {pages} {369} (\bibinfo {year} {1901})}\BibitemShut {NoStop}%
\bibitem [{\citenamefont {Crenshaw}(1996)}]{crenshaw_new_1996}%
  \BibitemOpen
  \bibfield  {author} {\bibinfo {author} {\bibfnamefont {H.~C.}\ \bibnamefont {Crenshaw}},\ }\bibfield  {title} {\bibinfo {title} {A {New} {Look} at {Locomotion} in {Microorganisms}: {Rotating} and {Translating}},\ }\href@noop {} {\bibfield  {journal} {\bibinfo  {journal} {Am. Zool.}\ }\textbf {\bibinfo {volume} {36}},\ \bibinfo {pages} {608} (\bibinfo {year} {1996})}\BibitemShut {NoStop}%
\bibitem [{\citenamefont {Berg}\ and\ \citenamefont {Turner}(1990)}]{berg_chemotaxis_1990}%
  \BibitemOpen
  \bibfield  {author} {\bibinfo {author} {\bibfnamefont {H.~C.}\ \bibnamefont {Berg}}\ and\ \bibinfo {author} {\bibfnamefont {L.}~\bibnamefont {Turner}},\ }\bibfield  {title} {\bibinfo {title} {Chemotaxis of bacteria in glass capillary arrays. {Escherichia} coli, motility, microchannel plate, and light scattering},\ }\href@noop {} {\bibfield  {journal} {\bibinfo  {journal} {Biophys. J.}\ }\textbf {\bibinfo {volume} {58}},\ \bibinfo {pages} {919} (\bibinfo {year} {1990})}\BibitemShut {NoStop}%
\bibitem [{\citenamefont {Thar}\ and\ \citenamefont {Fenchel}(2001)}]{thar_true_2001}%
  \BibitemOpen
  \bibfield  {author} {\bibinfo {author} {\bibfnamefont {R.}~\bibnamefont {Thar}}\ and\ \bibinfo {author} {\bibfnamefont {T.}~\bibnamefont {Fenchel}},\ }\bibfield  {title} {\bibinfo {title} {True {Chemotaxis} in {Oxygen} {Gradients} of the {Sulfur}-{Oxidizing} {Bacterium} {Thiovulum} majus},\ }\href@noop {} {\bibfield  {journal} {\bibinfo  {journal} {Appl. Environ. Microbiol.}\ }\textbf {\bibinfo {volume} {67}},\ \bibinfo {pages} {3299} (\bibinfo {year} {2001})}\BibitemShut {NoStop}%
\bibitem [{\citenamefont {Fenchel}\ and\ \citenamefont {Blackburn}(1999)}]{fenchel_motile_1999}%
  \BibitemOpen
  \bibfield  {author} {\bibinfo {author} {\bibfnamefont {T.}~\bibnamefont {Fenchel}}\ and\ \bibinfo {author} {\bibfnamefont {N.}~\bibnamefont {Blackburn}},\ }\bibfield  {title} {\bibinfo {title} {Motile {Chemosensory} {Behaviour} of {Phagotrophic} {Protists}: {Mechanisms} for and {Efficiency} in {Congregating} at {Food} {Patches}},\ }\href@noop {} {\bibfield  {journal} {\bibinfo  {journal} {Protist}\ }\textbf {\bibinfo {volume} {150}},\ \bibinfo {pages} {325} (\bibinfo {year} {1999})}\BibitemShut {NoStop}%
\bibitem [{\citenamefont {Jékely}\ \emph {et~al.}(2008)\citenamefont {Jékely}, \citenamefont {Colombelli}, \citenamefont {Hausen}, \citenamefont {Guy}, \citenamefont {Stelzer}, \citenamefont {Nédélec},\ and\ \citenamefont {Arendt}}]{jekely_mechanism_2008}%
  \BibitemOpen
  \bibfield  {author} {\bibinfo {author} {\bibfnamefont {G.}~\bibnamefont {Jékely}}, \bibinfo {author} {\bibfnamefont {J.}~\bibnamefont {Colombelli}}, \bibinfo {author} {\bibfnamefont {H.}~\bibnamefont {Hausen}}, \bibinfo {author} {\bibfnamefont {K.}~\bibnamefont {Guy}}, \bibinfo {author} {\bibfnamefont {E.}~\bibnamefont {Stelzer}}, \bibinfo {author} {\bibfnamefont {F.}~\bibnamefont {Nédélec}},\ and\ \bibinfo {author} {\bibfnamefont {D.}~\bibnamefont {Arendt}},\ }\bibfield  {title} {{\selectlanguage {en}\bibinfo {title} {Mechanism of phototaxis in marine zooplankton}},\ }\href@noop {} {\bibfield  {journal} {\bibinfo  {journal} {Nature}\ }\textbf {\bibinfo {volume} {456}},\ \bibinfo {pages} {395} (\bibinfo {year} {2008})}\BibitemShut {NoStop}%
\bibitem [{\citenamefont {Leptos}\ \emph {et~al.}(2023)\citenamefont {Leptos}, \citenamefont {Chioccioli}, \citenamefont {Furlan}, \citenamefont {Pesci},\ and\ \citenamefont {Goldstein}}]{leptos2023phototaxis}%
  \BibitemOpen
  \bibfield  {author} {\bibinfo {author} {\bibfnamefont {K.~C.}\ \bibnamefont {Leptos}}, \bibinfo {author} {\bibfnamefont {M.}~\bibnamefont {Chioccioli}}, \bibinfo {author} {\bibfnamefont {S.}~\bibnamefont {Furlan}}, \bibinfo {author} {\bibfnamefont {A.~I.}\ \bibnamefont {Pesci}},\ and\ \bibinfo {author} {\bibfnamefont {R.~E.}\ \bibnamefont {Goldstein}},\ }\bibfield  {title} {\bibinfo {title} {Phototaxis of chlamydomonas arises from a tuned adaptive photoresponse shared with multicellular volvocine green algae},\ }\href@noop {} {\bibfield  {journal} {\bibinfo  {journal} {Phys. Rev. E}\ }\textbf {\bibinfo {volume} {107}},\ \bibinfo {pages} {014404} (\bibinfo {year} {2023})}\BibitemShut {NoStop}%
\bibitem [{\citenamefont {Ripp}\ \emph {et~al.}(2021)\citenamefont {Ripp}, \citenamefont {Kehrer}, \citenamefont {Smyrnakou}, \citenamefont {Tisch}, \citenamefont {Tavares}, \citenamefont {Amino}, \citenamefont {Ruiz~de Almodovar},\ and\ \citenamefont {Frischknecht}}]{ripp_malaria_2021}%
  \BibitemOpen
  \bibfield  {author} {\bibinfo {author} {\bibfnamefont {J.}~\bibnamefont {Ripp}}, \bibinfo {author} {\bibfnamefont {J.}~\bibnamefont {Kehrer}}, \bibinfo {author} {\bibfnamefont {X.}~\bibnamefont {Smyrnakou}}, \bibinfo {author} {\bibfnamefont {N.}~\bibnamefont {Tisch}}, \bibinfo {author} {\bibfnamefont {J.}~\bibnamefont {Tavares}}, \bibinfo {author} {\bibfnamefont {R.}~\bibnamefont {Amino}}, \bibinfo {author} {\bibfnamefont {C.}~\bibnamefont {Ruiz~de Almodovar}},\ and\ \bibinfo {author} {\bibfnamefont {F.}~\bibnamefont {Frischknecht}},\ }\bibfield  {title} {{\selectlanguage {en}\bibinfo {title} {Malaria parasites differentially sense environmental elasticity during transmission}},\ }\href@noop {} {\bibfield  {journal} {\bibinfo  {journal} {EMBO Mol. Med.}\ }\textbf {\bibinfo {volume} {13}},\ \bibinfo {pages} {e13933} (\bibinfo {year} {2021})}\BibitemShut {NoStop}%
\bibitem [{\citenamefont {Liu}\ \emph {et~al.}(2024)\citenamefont {Liu}, \citenamefont {Li}, \citenamefont {Anantha}, \citenamefont {Thanakornsombut}, \citenamefont {Wu}, \citenamefont {Chen}, \citenamefont {Tsuchiya}, \citenamefont {Tripathi}, \citenamefont {Chen},\ and\ \citenamefont {Barman}}]{liu_plasmodium_2024}%
  \BibitemOpen
  \bibfield  {author} {\bibinfo {author} {\bibfnamefont {Z.}~\bibnamefont {Liu}}, \bibinfo {author} {\bibfnamefont {S.}~\bibnamefont {Li}}, \bibinfo {author} {\bibfnamefont {P.}~\bibnamefont {Anantha}}, \bibinfo {author} {\bibfnamefont {T.}~\bibnamefont {Thanakornsombut}}, \bibinfo {author} {\bibfnamefont {L.}~\bibnamefont {Wu}}, \bibinfo {author} {\bibfnamefont {J.}~\bibnamefont {Chen}}, \bibinfo {author} {\bibfnamefont {R.}~\bibnamefont {Tsuchiya}}, \bibinfo {author} {\bibfnamefont {A.~K.}\ \bibnamefont {Tripathi}}, \bibinfo {author} {\bibfnamefont {Y.}~\bibnamefont {Chen}},\ and\ \bibinfo {author} {\bibfnamefont {I.}~\bibnamefont {Barman}},\ }\bibfield  {title} {{\selectlanguage {English}\bibinfo {title} {Plasmodium sporozoite shows distinct motility patterns in responses to three-dimensional environments}},\ }\href@noop {} {\bibfield  {journal} {\bibinfo  {journal} {iScience}\ }\textbf {\bibinfo {volume} {27}} (\bibinfo {year} {2024})}\BibitemShut {NoStop}%
\bibitem [{\citenamefont {Lettermann}\ \emph {et~al.}(2024)\citenamefont {Lettermann}, \citenamefont {Ziebert},\ and\ \citenamefont {Schwarz}}]{lettermann_geometrical_2024}%
  \BibitemOpen
  \bibfield  {author} {\bibinfo {author} {\bibfnamefont {L.}~\bibnamefont {Lettermann}}, \bibinfo {author} {\bibfnamefont {F.}~\bibnamefont {Ziebert}},\ and\ \bibinfo {author} {\bibfnamefont {U.~S.}\ \bibnamefont {Schwarz}},\ }\bibfield  {title} {\bibinfo {title} {A geometrical theory of gliding motility based on cell shape and surface flow},\ }\href@noop {} {\bibfield  {journal} {\bibinfo  {journal} {Proc. Natl. Acad. Sci.}\ }\textbf {\bibinfo {volume} {121}},\ \bibinfo {pages} {e2410708121} (\bibinfo {year} {2024})}\BibitemShut {NoStop}%
\bibitem [{\citenamefont {Berg}\ and\ \citenamefont {Purcell}(1977)}]{berg1977physics}%
  \BibitemOpen
  \bibfield  {author} {\bibinfo {author} {\bibfnamefont {H.~C.}\ \bibnamefont {Berg}}\ and\ \bibinfo {author} {\bibfnamefont {E.~M.}\ \bibnamefont {Purcell}},\ }\bibfield  {title} {\bibinfo {title} {Physics of chemoreception},\ }\href@noop {} {\bibfield  {journal} {\bibinfo  {journal} {Biophys. J.}\ }\textbf {\bibinfo {volume} {20}},\ \bibinfo {pages} {193} (\bibinfo {year} {1977})}\BibitemShut {NoStop}%
\bibitem [{\citenamefont {Berg}(1993)}]{berg1993random}%
  \BibitemOpen
  \bibfield  {author} {\bibinfo {author} {\bibfnamefont {H.~C.}\ \bibnamefont {Berg}},\ }\href@noop {} {\emph {\bibinfo {title} {Random walks in biology}}}\ (\bibinfo  {publisher} {Princeton University Press},\ \bibinfo {year} {1993})\BibitemShut {NoStop}%
\bibitem [{\citenamefont {Pohl}\ \emph {et~al.}(2017)\citenamefont {Pohl}, \citenamefont {Hintsche}, \citenamefont {Alirezaeizanjani}, \citenamefont {Seyrich}, \citenamefont {Beta},\ and\ \citenamefont {Stark}}]{pohl2017inferring}%
  \BibitemOpen
  \bibfield  {author} {\bibinfo {author} {\bibfnamefont {O.}~\bibnamefont {Pohl}}, \bibinfo {author} {\bibfnamefont {M.}~\bibnamefont {Hintsche}}, \bibinfo {author} {\bibfnamefont {Z.}~\bibnamefont {Alirezaeizanjani}}, \bibinfo {author} {\bibfnamefont {M.}~\bibnamefont {Seyrich}}, \bibinfo {author} {\bibfnamefont {C.}~\bibnamefont {Beta}},\ and\ \bibinfo {author} {\bibfnamefont {H.}~\bibnamefont {Stark}},\ }\bibfield  {title} {\bibinfo {title} {Inferring the chemotactic strategy of p. putida and e. coli using modified kramers-moyal coefficients},\ }\href@noop {} {\bibfield  {journal} {\bibinfo  {journal} {PLoS Comput. Biol.}\ }\textbf {\bibinfo {volume} {13}},\ \bibinfo {pages} {e1005329} (\bibinfo {year} {2017})}\BibitemShut {NoStop}%
\bibitem [{\citenamefont {Romanczuk}\ \emph {et~al.}(2012)\citenamefont {Romanczuk}, \citenamefont {Bär}, \citenamefont {Ebeling}, \citenamefont {Lindner},\ and\ \citenamefont {Schimansky-Geier}}]{romanczuk_active_2012}%
  \BibitemOpen
  \bibfield  {author} {\bibinfo {author} {\bibfnamefont {P.}~\bibnamefont {Romanczuk}}, \bibinfo {author} {\bibfnamefont {M.}~\bibnamefont {Bär}}, \bibinfo {author} {\bibfnamefont {W.}~\bibnamefont {Ebeling}}, \bibinfo {author} {\bibfnamefont {B.}~\bibnamefont {Lindner}},\ and\ \bibinfo {author} {\bibfnamefont {L.}~\bibnamefont {Schimansky-Geier}},\ }\bibfield  {title} {{\selectlanguage {en}\bibinfo {title} {Active {Brownian} particles}},\ }\href@noop {} {\bibfield  {journal} {\bibinfo  {journal} {Eur. Phys. J. Spec. Top.}\ }\textbf {\bibinfo {volume} {202}},\ \bibinfo {pages} {1} (\bibinfo {year} {2012})}\BibitemShut {NoStop}%
\bibitem [{\citenamefont {Bechinger}\ \emph {et~al.}(2016)\citenamefont {Bechinger}, \citenamefont {Di~Leonardo}, \citenamefont {Löwen}, \citenamefont {Reichhardt}, \citenamefont {Volpe},\ and\ \citenamefont {Volpe}}]{bechinger_active_2016}%
  \BibitemOpen
  \bibfield  {author} {\bibinfo {author} {\bibfnamefont {C.}~\bibnamefont {Bechinger}}, \bibinfo {author} {\bibfnamefont {R.}~\bibnamefont {Di~Leonardo}}, \bibinfo {author} {\bibfnamefont {H.}~\bibnamefont {Löwen}}, \bibinfo {author} {\bibfnamefont {C.}~\bibnamefont {Reichhardt}}, \bibinfo {author} {\bibfnamefont {G.}~\bibnamefont {Volpe}},\ and\ \bibinfo {author} {\bibfnamefont {G.}~\bibnamefont {Volpe}},\ }\bibfield  {title} {{\selectlanguage {en}\bibinfo {title} {Active {Particles} in {Complex} and {Crowded} {Environments}}},\ }\href@noop {} {\bibfield  {journal} {\bibinfo  {journal} {Rev. Mod. Phys.}\ }\textbf {\bibinfo {volume} {88}},\ \bibinfo {pages} {045006} (\bibinfo {year} {2016})}\BibitemShut {NoStop}%
\bibitem [{\citenamefont {Zöttl}\ and\ \citenamefont {Stark}(2016)}]{zottl_emergent_2016}%
  \BibitemOpen
  \bibfield  {author} {\bibinfo {author} {\bibfnamefont {A.}~\bibnamefont {Zöttl}}\ and\ \bibinfo {author} {\bibfnamefont {H.}~\bibnamefont {Stark}},\ }\bibfield  {title} {{\selectlanguage {en}\bibinfo {title} {Emergent behavior in active colloids}},\ }\href@noop {} {\bibfield  {journal} {\bibinfo  {journal} {J. Phys. Condens. Matter.}\ }\textbf {\bibinfo {volume} {28}},\ \bibinfo {pages} {253001} (\bibinfo {year} {2016})}\BibitemShut {NoStop}%
\bibitem [{\citenamefont {Liebchen}\ and\ \citenamefont {Levis}(2022)}]{liebchen2022chiral}%
  \BibitemOpen
  \bibfield  {author} {\bibinfo {author} {\bibfnamefont {B.}~\bibnamefont {Liebchen}}\ and\ \bibinfo {author} {\bibfnamefont {D.}~\bibnamefont {Levis}},\ }\bibfield  {title} {\bibinfo {title} {Chiral active matter},\ }\href@noop {} {\bibfield  {journal} {\bibinfo  {journal} {Europhys. Lett.}\ }\textbf {\bibinfo {volume} {139}},\ \bibinfo {pages} {67001} (\bibinfo {year} {2022})}\BibitemShut {NoStop}%
\bibitem [{\citenamefont {Howse}\ \emph {et~al.}(2007)\citenamefont {Howse}, \citenamefont {Jones}, \citenamefont {Ryan}, \citenamefont {Gough}, \citenamefont {Vafabakhsh},\ and\ \citenamefont {Golestanian}}]{howse_self-motile_2007}%
  \BibitemOpen
  \bibfield  {author} {\bibinfo {author} {\bibfnamefont {J.~R.}\ \bibnamefont {Howse}}, \bibinfo {author} {\bibfnamefont {R.~A.~L.}\ \bibnamefont {Jones}}, \bibinfo {author} {\bibfnamefont {A.~J.}\ \bibnamefont {Ryan}}, \bibinfo {author} {\bibfnamefont {T.}~\bibnamefont {Gough}}, \bibinfo {author} {\bibfnamefont {R.}~\bibnamefont {Vafabakhsh}},\ and\ \bibinfo {author} {\bibfnamefont {R.}~\bibnamefont {Golestanian}},\ }\bibfield  {title} {\bibinfo {title} {Self-{Motile} {Colloidal} {Particles}: {From} {Directed} {Propulsion} to {Random} {Walk}},\ }\href@noop {} {\bibfield  {journal} {\bibinfo  {journal} {Phys. Rev. Lett.}\ }\textbf {\bibinfo {volume} {99}},\ \bibinfo {pages} {048102} (\bibinfo {year} {2007})}\BibitemShut {NoStop}%
\bibitem [{\citenamefont {Debnath}\ \emph {et~al.}(2016)\citenamefont {Debnath}, \citenamefont {K.~Ghosh}, \citenamefont {Li}, \citenamefont {Marchesoni},\ and\ \citenamefont {Li}}]{debnath_diffusion_2016}%
  \BibitemOpen
  \bibfield  {author} {\bibinfo {author} {\bibfnamefont {D.}~\bibnamefont {Debnath}}, \bibinfo {author} {\bibfnamefont {P.}~\bibnamefont {K.~Ghosh}}, \bibinfo {author} {\bibfnamefont {Y.}~\bibnamefont {Li}}, \bibinfo {author} {\bibfnamefont {F.}~\bibnamefont {Marchesoni}},\ and\ \bibinfo {author} {\bibfnamefont {B.}~\bibnamefont {Li}},\ }\bibfield  {title} {{\selectlanguage {en}\bibinfo {title} {Diffusion of eccentric microswimmers}},\ }\href@noop {} {\bibfield  {journal} {\bibinfo  {journal} {Soft Matter}\ }\textbf {\bibinfo {volume} {12}},\ \bibinfo {pages} {2017} (\bibinfo {year} {2016})}\BibitemShut {NoStop}%
\bibitem [{\citenamefont {Gomez-Solano}\ \emph {et~al.}(2016)\citenamefont {Gomez-Solano}, \citenamefont {Blokhuis},\ and\ \citenamefont {Bechinger}}]{gomez-solano_dynamics_2016}%
  \BibitemOpen
  \bibfield  {author} {\bibinfo {author} {\bibfnamefont {J.~R.}\ \bibnamefont {Gomez-Solano}}, \bibinfo {author} {\bibfnamefont {A.}~\bibnamefont {Blokhuis}},\ and\ \bibinfo {author} {\bibfnamefont {C.}~\bibnamefont {Bechinger}},\ }\bibfield  {title} {\bibinfo {title} {Dynamics of {Self}-{Propelled} {Janus} {Particles} in {Viscoelastic} {Fluids}},\ }\href@noop {} {\bibfield  {journal} {\bibinfo  {journal} {Phys. Rev. Lett.}\ }\textbf {\bibinfo {volume} {116}},\ \bibinfo {pages} {138301} (\bibinfo {year} {2016})}\BibitemShut {NoStop}%
\bibitem [{\citenamefont {Sevilla}(2016)}]{sevilla2016diffusion}%
  \BibitemOpen
  \bibfield  {author} {\bibinfo {author} {\bibfnamefont {F.~J.}\ \bibnamefont {Sevilla}},\ }\bibfield  {title} {\bibinfo {title} {Diffusion of active chiral particles},\ }\href@noop {} {\bibfield  {journal} {\bibinfo  {journal} {Phys. Rev. E}\ }\textbf {\bibinfo {volume} {94}},\ \bibinfo {pages} {062120} (\bibinfo {year} {2016})}\BibitemShut {NoStop}%
\bibitem [{\citenamefont {Gomez-Solano}\ and\ \citenamefont {Sevilla}(2020)}]{gomez-solano_active_2020}%
  \BibitemOpen
  \bibfield  {author} {\bibinfo {author} {\bibfnamefont {J.~R.}\ \bibnamefont {Gomez-Solano}}\ and\ \bibinfo {author} {\bibfnamefont {F.~J.}\ \bibnamefont {Sevilla}},\ }\bibfield  {title} {{\selectlanguage {en}\bibinfo {title} {Active particles with fractional rotational {Brownian} motion}},\ }\href@noop {} {\bibfield  {journal} {\bibinfo  {journal} {J. Stat. Mech. Theor. Exp.}\ }\textbf {\bibinfo {volume} {2020}},\ \bibinfo {pages} {063213} (\bibinfo {year} {2020})}\BibitemShut {NoStop}%
\bibitem [{\citenamefont {Sprenger}\ \emph {et~al.}(2023)\citenamefont {Sprenger}, \citenamefont {Caprini}, \citenamefont {Löwen},\ and\ \citenamefont {Wittmann}}]{sprenger_dynamics_2023}%
  \BibitemOpen
  \bibfield  {author} {\bibinfo {author} {\bibfnamefont {A.~R.}\ \bibnamefont {Sprenger}}, \bibinfo {author} {\bibfnamefont {L.}~\bibnamefont {Caprini}}, \bibinfo {author} {\bibfnamefont {H.}~\bibnamefont {Löwen}},\ and\ \bibinfo {author} {\bibfnamefont {R.}~\bibnamefont {Wittmann}},\ }\bibfield  {title} {{\selectlanguage {en}\bibinfo {title} {Dynamics of active particles with translational and rotational inertia}},\ }\href@noop {} {\bibfield  {journal} {\bibinfo  {journal} {J. Phys. Condens. Matter.}\ }\textbf {\bibinfo {volume} {35}},\ \bibinfo {pages} {305101} (\bibinfo {year} {2023})}\BibitemShut {NoStop}%
\bibitem [{\citenamefont {van Teeffelen}\ and\ \citenamefont {Löwen}(2008)}]{van_teeffelen_dynamics_2008}%
  \BibitemOpen
  \bibfield  {author} {\bibinfo {author} {\bibfnamefont {S.}~\bibnamefont {van Teeffelen}}\ and\ \bibinfo {author} {\bibfnamefont {H.}~\bibnamefont {Löwen}},\ }\bibfield  {title} {\bibinfo {title} {Dynamics of a {Brownian} circle swimmer},\ }\href@noop {} {\bibfield  {journal} {\bibinfo  {journal} {Phys. Rev. E}\ }\textbf {\bibinfo {volume} {78}},\ \bibinfo {pages} {020101} (\bibinfo {year} {2008})}\BibitemShut {NoStop}%
\bibitem [{\citenamefont {Ledesma-Aguilar}\ \emph {et~al.}(2012)\citenamefont {Ledesma-Aguilar}, \citenamefont {Löwen},\ and\ \citenamefont {Yeomans}}]{ledesma-aguilar_circle_2012}%
  \BibitemOpen
  \bibfield  {author} {\bibinfo {author} {\bibfnamefont {R.}~\bibnamefont {Ledesma-Aguilar}}, \bibinfo {author} {\bibfnamefont {H.}~\bibnamefont {Löwen}},\ and\ \bibinfo {author} {\bibfnamefont {J.~M.}\ \bibnamefont {Yeomans}},\ }\bibfield  {title} {{\selectlanguage {en}\bibinfo {title} {A circle swimmer at low {Reynolds} number}},\ }\href@noop {} {\bibfield  {journal} {\bibinfo  {journal} {Eur. Phys. J. E}\ }\textbf {\bibinfo {volume} {35}},\ \bibinfo {pages} {70} (\bibinfo {year} {2012})}\BibitemShut {NoStop}%
\bibitem [{\citenamefont {Kümmel}\ \emph {et~al.}(2013)\citenamefont {Kümmel}, \citenamefont {ten Hagen}, \citenamefont {Wittkowski}, \citenamefont {Buttinoni}, \citenamefont {Eichhorn}, \citenamefont {Volpe}, \citenamefont {Löwen},\ and\ \citenamefont {Bechinger}}]{kummel_circular_2013}%
  \BibitemOpen
  \bibfield  {author} {\bibinfo {author} {\bibfnamefont {F.}~\bibnamefont {Kümmel}}, \bibinfo {author} {\bibfnamefont {B.}~\bibnamefont {ten Hagen}}, \bibinfo {author} {\bibfnamefont {R.}~\bibnamefont {Wittkowski}}, \bibinfo {author} {\bibfnamefont {I.}~\bibnamefont {Buttinoni}}, \bibinfo {author} {\bibfnamefont {R.}~\bibnamefont {Eichhorn}}, \bibinfo {author} {\bibfnamefont {G.}~\bibnamefont {Volpe}}, \bibinfo {author} {\bibfnamefont {H.}~\bibnamefont {Löwen}},\ and\ \bibinfo {author} {\bibfnamefont {C.}~\bibnamefont {Bechinger}},\ }\bibfield  {title} {\bibinfo {title} {Circular {Motion} of {Asymmetric} {Self}-{Propelling} {Particles}},\ }\href@noop {} {\bibfield  {journal} {\bibinfo  {journal} {Phys. Rev. Lett.}\ }\textbf {\bibinfo {volume} {110}},\ \bibinfo {pages} {198302} (\bibinfo {year} {2013})}\BibitemShut {NoStop}%
\bibitem [{\citenamefont {Marine}\ \emph {et~al.}(2013)\citenamefont {Marine}, \citenamefont {Wheat}, \citenamefont {Ault},\ and\ \citenamefont {Posner}}]{marine_diffusive_2013}%
  \BibitemOpen
  \bibfield  {author} {\bibinfo {author} {\bibfnamefont {N.~A.}\ \bibnamefont {Marine}}, \bibinfo {author} {\bibfnamefont {P.~M.}\ \bibnamefont {Wheat}}, \bibinfo {author} {\bibfnamefont {J.}~\bibnamefont {Ault}},\ and\ \bibinfo {author} {\bibfnamefont {J.~D.}\ \bibnamefont {Posner}},\ }\bibfield  {title} {\bibinfo {title} {Diffusive behaviors of circle-swimming motors},\ }\href@noop {} {\bibfield  {journal} {\bibinfo  {journal} {Phys. Rev. E}\ }\textbf {\bibinfo {volume} {87}},\ \bibinfo {pages} {052305} (\bibinfo {year} {2013})}\BibitemShut {NoStop}%
\bibitem [{\citenamefont {Löwen}(2016)}]{lowen_chirality_2016}%
  \BibitemOpen
  \bibfield  {author} {\bibinfo {author} {\bibfnamefont {H.}~\bibnamefont {Löwen}},\ }\bibfield  {title} {{\selectlanguage {en}\bibinfo {title} {Chirality in microswimmer motion: {From} circle swimmers to active turbulence}},\ }\href@noop {} {\bibfield  {journal} {\bibinfo  {journal} {Eur. Phys. J. Spec. Top.}\ }\textbf {\bibinfo {volume} {225}},\ \bibinfo {pages} {2319} (\bibinfo {year} {2016})}\BibitemShut {NoStop}%
\bibitem [{\citenamefont {Nourhani}\ \emph {et~al.}(2016)\citenamefont {Nourhani}, \citenamefont {Ebbens}, \citenamefont {Gibbs},\ and\ \citenamefont {Lammert}}]{nourhani2016spiral}%
  \BibitemOpen
  \bibfield  {author} {\bibinfo {author} {\bibfnamefont {A.}~\bibnamefont {Nourhani}}, \bibinfo {author} {\bibfnamefont {S.~J.}\ \bibnamefont {Ebbens}}, \bibinfo {author} {\bibfnamefont {J.~G.}\ \bibnamefont {Gibbs}},\ and\ \bibinfo {author} {\bibfnamefont {P.~E.}\ \bibnamefont {Lammert}},\ }\bibfield  {title} {\bibinfo {title} {Spiral diffusion of rotating self-propellers with stochastic perturbation},\ }\href@noop {} {\bibfield  {journal} {\bibinfo  {journal} {Phys. Rev. E}\ }\textbf {\bibinfo {volume} {94}},\ \bibinfo {pages} {030601} (\bibinfo {year} {2016})}\BibitemShut {NoStop}%
\bibitem [{\citenamefont {Caprini}\ \emph {et~al.}(2023)\citenamefont {Caprini}, \citenamefont {Löwen},\ and\ \citenamefont {Marconi}}]{caprini_chiral_2023}%
  \BibitemOpen
  \bibfield  {author} {\bibinfo {author} {\bibfnamefont {L.}~\bibnamefont {Caprini}}, \bibinfo {author} {\bibfnamefont {H.}~\bibnamefont {Löwen}},\ and\ \bibinfo {author} {\bibfnamefont {U.~M.~B.}\ \bibnamefont {Marconi}},\ }\bibfield  {title} {{\selectlanguage {en}\bibinfo {title} {Chiral active matter in external potentials}},\ }\href@noop {} {\bibfield  {journal} {\bibinfo  {journal} {Soft Matter}\ }\textbf {\bibinfo {volume} {19}},\ \bibinfo {pages} {6234} (\bibinfo {year} {2023})}\BibitemShut {NoStop}%
\bibitem [{\citenamefont {Lindner}(2010)}]{lindner_diffusion_2010}%
  \BibitemOpen
  \bibfield  {author} {\bibinfo {author} {\bibfnamefont {B.}~\bibnamefont {Lindner}},\ }\bibfield  {title} {{\selectlanguage {en}\bibinfo {title} {Diffusion of particles subject to nonlinear friction and a colored noise}},\ }\href@noop {} {\bibfield  {journal} {\bibinfo  {journal} {New J. Phys.}\ }\textbf {\bibinfo {volume} {12}},\ \bibinfo {pages} {063026} (\bibinfo {year} {2010})}\BibitemShut {NoStop}%
\bibitem [{\citenamefont {Weber}\ \emph {et~al.}(2011)\citenamefont {Weber}, \citenamefont {Radtke}, \citenamefont {Schimansky-Geier},\ and\ \citenamefont {Hänggi}}]{weber_active_2011}%
  \BibitemOpen
  \bibfield  {author} {\bibinfo {author} {\bibfnamefont {C.}~\bibnamefont {Weber}}, \bibinfo {author} {\bibfnamefont {P.~K.}\ \bibnamefont {Radtke}}, \bibinfo {author} {\bibfnamefont {L.}~\bibnamefont {Schimansky-Geier}},\ and\ \bibinfo {author} {\bibfnamefont {P.}~\bibnamefont {Hänggi}},\ }\bibfield  {title} {\bibinfo {title} {Active motion assisted by correlated stochastic torques},\ }\href@noop {} {\bibfield  {journal} {\bibinfo  {journal} {Phys. Rev. E}\ }\textbf {\bibinfo {volume} {84}},\ \bibinfo {pages} {011132} (\bibinfo {year} {2011})}\BibitemShut {NoStop}%
\bibitem [{\citenamefont {Ghosh}\ \emph {et~al.}(2015)\citenamefont {Ghosh}, \citenamefont {Li}, \citenamefont {Marchegiani},\ and\ \citenamefont {Marchesoni}}]{ghosh_communication_2015}%
  \BibitemOpen
  \bibfield  {author} {\bibinfo {author} {\bibfnamefont {P.~K.}\ \bibnamefont {Ghosh}}, \bibinfo {author} {\bibfnamefont {Y.}~\bibnamefont {Li}}, \bibinfo {author} {\bibfnamefont {G.}~\bibnamefont {Marchegiani}},\ and\ \bibinfo {author} {\bibfnamefont {F.}~\bibnamefont {Marchesoni}},\ }\bibfield  {title} {\bibinfo {title} {Communication: {Memory} effects and active {Brownian} diffusion},\ }\href@noop {} {\bibfield  {journal} {\bibinfo  {journal} {J. Chem. Phys.}\ }\textbf {\bibinfo {volume} {143}},\ \bibinfo {pages} {211101} (\bibinfo {year} {2015})}\BibitemShut {NoStop}%
\bibitem [{\citenamefont {Narinder}\ \emph {et~al.}(2018)\citenamefont {Narinder}, \citenamefont {Bechinger},\ and\ \citenamefont {Gomez-Solano}}]{narinder_memory-induced_2018}%
  \BibitemOpen
  \bibfield  {author} {\bibinfo {author} {\bibfnamefont {N.}~\bibnamefont {Narinder}}, \bibinfo {author} {\bibfnamefont {C.}~\bibnamefont {Bechinger}},\ and\ \bibinfo {author} {\bibfnamefont {J.~R.}\ \bibnamefont {Gomez-Solano}},\ }\bibfield  {title} {\bibinfo {title} {Memory-{Induced} {Transition} from a {Persistent} {Random} {Walk} to {Circular} {Motion} for {Achiral} {Microswimmers}},\ }\href@noop {} {\bibfield  {journal} {\bibinfo  {journal} {Phys. Rev. Lett.}\ }\textbf {\bibinfo {volume} {121}},\ \bibinfo {pages} {078003} (\bibinfo {year} {2018})}\BibitemShut {NoStop}%
\bibitem [{\citenamefont {Bayati}\ and\ \citenamefont {Nourhani}(2022)}]{bayati_memory_2022}%
  \BibitemOpen
  \bibfield  {author} {\bibinfo {author} {\bibfnamefont {P.}~\bibnamefont {Bayati}}\ and\ \bibinfo {author} {\bibfnamefont {A.}~\bibnamefont {Nourhani}},\ }\bibfield  {title} {\bibinfo {title} {Memory effects in spiral diffusion of rotary self-propellers},\ }\href@noop {} {\bibfield  {journal} {\bibinfo  {journal} {Phys. Rev. E}\ }\textbf {\bibinfo {volume} {105}},\ \bibinfo {pages} {024606} (\bibinfo {year} {2022})}\BibitemShut {NoStop}%
\bibitem [{\citenamefont {Sprenger}\ \emph {et~al.}(2021)\citenamefont {Sprenger}, \citenamefont {Jahanshahi}, \citenamefont {Ivlev},\ and\ \citenamefont {Löwen}}]{sprenger_time-dependent_2021}%
  \BibitemOpen
  \bibfield  {author} {\bibinfo {author} {\bibfnamefont {A.~R.}\ \bibnamefont {Sprenger}}, \bibinfo {author} {\bibfnamefont {S.}~\bibnamefont {Jahanshahi}}, \bibinfo {author} {\bibfnamefont {A.~V.}\ \bibnamefont {Ivlev}},\ and\ \bibinfo {author} {\bibfnamefont {H.}~\bibnamefont {Löwen}},\ }\bibfield  {title} {\bibinfo {title} {Time-dependent inertia of self-propelled particles: {The} {Langevin} rocket},\ }\href@noop {} {\bibfield  {journal} {\bibinfo  {journal} {Phys. Rev. E}\ }\textbf {\bibinfo {volume} {103}},\ \bibinfo {pages} {042601} (\bibinfo {year} {2021})}\BibitemShut {NoStop}%
\bibitem [{\citenamefont {Rutenberg}\ \emph {et~al.}(2003)\citenamefont {Rutenberg}, \citenamefont {Richardson},\ and\ \citenamefont {Montgomery}}]{rutenberg_diffusion_2003}%
  \BibitemOpen
  \bibfield  {author} {\bibinfo {author} {\bibfnamefont {A.~D.}\ \bibnamefont {Rutenberg}}, \bibinfo {author} {\bibfnamefont {A.~J.}\ \bibnamefont {Richardson}},\ and\ \bibinfo {author} {\bibfnamefont {C.~J.}\ \bibnamefont {Montgomery}},\ }\bibfield  {title} {{\selectlanguage {en}\bibinfo {title} {Diffusion of {Asymmetric} {Swimmers}}},\ }\href@noop {} {\bibfield  {journal} {\bibinfo  {journal} {Phys. Rev. Lett.}\ }\textbf {\bibinfo {volume} {91}},\ \bibinfo {pages} {080601} (\bibinfo {year} {2003})}\BibitemShut {NoStop}%
\bibitem [{\citenamefont {Becker}\ and\ \citenamefont {Everaers}(2007)}]{becker_rigid_2007}%
  \BibitemOpen
  \bibfield  {author} {\bibinfo {author} {\bibfnamefont {N.~B.}\ \bibnamefont {Becker}}\ and\ \bibinfo {author} {\bibfnamefont {R.}~\bibnamefont {Everaers}},\ }\bibfield  {title} {\bibinfo {title} {From rigid base pairs to semiflexible polymers: {Coarse}-graining {DNA}},\ }\href@noop {} {\bibfield  {journal} {\bibinfo  {journal} {Phys. Rev. E}\ }\textbf {\bibinfo {volume} {76}},\ \bibinfo {pages} {021923} (\bibinfo {year} {2007})}\BibitemShut {NoStop}%
\bibitem [{\citenamefont {Friedrich}\ and\ \citenamefont {Jülicher}(2008)}]{friedrich_stochastic_2008}%
  \BibitemOpen
  \bibfield  {author} {\bibinfo {author} {\bibfnamefont {B.~M.}\ \bibnamefont {Friedrich}}\ and\ \bibinfo {author} {\bibfnamefont {F.}~\bibnamefont {Jülicher}},\ }\bibfield  {title} {{\selectlanguage {en}\bibinfo {title} {The stochastic dance of circling sperm cells: sperm chemotaxis in the plane}},\ }\href@noop {} {\bibfield  {journal} {\bibinfo  {journal} {New J. Phys.}\ }\textbf {\bibinfo {volume} {10}},\ \bibinfo {pages} {123025} (\bibinfo {year} {2008})}\BibitemShut {NoStop}%
\bibitem [{\citenamefont {Friedrich}\ and\ \citenamefont {Jülicher}(2009)}]{friedrich_steering_2009}%
  \BibitemOpen
  \bibfield  {author} {\bibinfo {author} {\bibfnamefont {B.~M.}\ \bibnamefont {Friedrich}}\ and\ \bibinfo {author} {\bibfnamefont {F.}~\bibnamefont {Jülicher}},\ }\bibfield  {title} {\bibinfo {title} {Steering {Chiral} {Swimmers} along {Noisy} {Helical} {Paths}},\ }\href@noop {} {\bibfield  {journal} {\bibinfo  {journal} {Phys. Rev. Lett.}\ }\textbf {\bibinfo {volume} {103}},\ \bibinfo {pages} {068102} (\bibinfo {year} {2009})}\BibitemShut {NoStop}%
\bibitem [{\citenamefont {Wittkowski}\ and\ \citenamefont {Löwen}(2012)}]{wittkowski_self-propelled_2012}%
  \BibitemOpen
  \bibfield  {author} {\bibinfo {author} {\bibfnamefont {R.}~\bibnamefont {Wittkowski}}\ and\ \bibinfo {author} {\bibfnamefont {H.}~\bibnamefont {Löwen}},\ }\bibfield  {title} {\bibinfo {title} {Self-propelled {Brownian} spinning top: {Dynamics} of a biaxial swimmer at low {Reynolds} numbers},\ }\href@noop {} {\bibfield  {journal} {\bibinfo  {journal} {Phys. Rev. E}\ }\textbf {\bibinfo {volume} {85}},\ \bibinfo {pages} {021406} (\bibinfo {year} {2012})}\BibitemShut {NoStop}%
\bibitem [{\citenamefont {Kirkegaard}\ \emph {et~al.}(2016)\citenamefont {Kirkegaard}, \citenamefont {Marron},\ and\ \citenamefont {Goldstein}}]{kirkegaard2016motility}%
  \BibitemOpen
  \bibfield  {author} {\bibinfo {author} {\bibfnamefont {J.~B.}\ \bibnamefont {Kirkegaard}}, \bibinfo {author} {\bibfnamefont {A.~O.}\ \bibnamefont {Marron}},\ and\ \bibinfo {author} {\bibfnamefont {R.~E.}\ \bibnamefont {Goldstein}},\ }\bibfield  {title} {\bibinfo {title} {Motility of colonial choanoflagellates and the statistics of aggregate random walkers},\ }\href@noop {} {\bibfield  {journal} {\bibinfo  {journal} {Phys. Rev. Lett.}\ }\textbf {\bibinfo {volume} {116}},\ \bibinfo {pages} {038102} (\bibinfo {year} {2016})}\BibitemShut {NoStop}%
\bibitem [{\citenamefont {Ren}\ and\ \citenamefont {Bloomfield-Gad{\^e}lha}(2025)}]{ren2025swimming}%
  \BibitemOpen
  \bibfield  {author} {\bibinfo {author} {\bibfnamefont {X.}~\bibnamefont {Ren}}\ and\ \bibinfo {author} {\bibfnamefont {H.}~\bibnamefont {Bloomfield-Gad{\^e}lha}},\ }\bibfield  {title} {\bibinfo {title} {Swimming by spinning: Spinning-top type rotations regularize sperm swimming into persistently progressive paths in 3d},\ }\href@noop {} {\bibfield  {journal} {\bibinfo  {journal} {Adv. Sci.}\ }\textbf {\bibinfo {volume} {12}},\ \bibinfo {pages} {2406143} (\bibinfo {year} {2025})}\BibitemShut {NoStop}%
\bibitem [{\citenamefont {Szamel}(2014)}]{szamel_self-propelled_2014}%
  \BibitemOpen
  \bibfield  {author} {\bibinfo {author} {\bibfnamefont {G.}~\bibnamefont {Szamel}},\ }\bibfield  {title} {\bibinfo {title} {Self-propelled particle in an external potential: {Existence} of an effective temperature},\ }\href@noop {} {\bibfield  {journal} {\bibinfo  {journal} {Phys. Rev. E}\ }\textbf {\bibinfo {volume} {90}},\ \bibinfo {pages} {012111} (\bibinfo {year} {2014})}\BibitemShut {NoStop}%
\bibitem [{\citenamefont {Sevilla}\ \emph {et~al.}(2019)\citenamefont {Sevilla}, \citenamefont {Rodríguez},\ and\ \citenamefont {Gomez-Solano}}]{sevilla_generalized_2019}%
  \BibitemOpen
  \bibfield  {author} {\bibinfo {author} {\bibfnamefont {F.~J.}\ \bibnamefont {Sevilla}}, \bibinfo {author} {\bibfnamefont {R.~F.}\ \bibnamefont {Rodríguez}},\ and\ \bibinfo {author} {\bibfnamefont {J.~R.}\ \bibnamefont {Gomez-Solano}},\ }\bibfield  {title} {\bibinfo {title} {Generalized {Ornstein}-{Uhlenbeck} model for active motion},\ }\href@noop {} {\bibfield  {journal} {\bibinfo  {journal} {Phys. Rev. E}\ }\textbf {\bibinfo {volume} {100}},\ \bibinfo {pages} {032123} (\bibinfo {year} {2019})}\BibitemShut {NoStop}%
\bibitem [{\citenamefont {Woillez}\ \emph {et~al.}(2020)\citenamefont {Woillez}, \citenamefont {Kafri},\ and\ \citenamefont {Lecomte}}]{woillez_nonlocal_2020}%
  \BibitemOpen
  \bibfield  {author} {\bibinfo {author} {\bibfnamefont {E.}~\bibnamefont {Woillez}}, \bibinfo {author} {\bibfnamefont {Y.}~\bibnamefont {Kafri}},\ and\ \bibinfo {author} {\bibfnamefont {V.}~\bibnamefont {Lecomte}},\ }\bibfield  {title} {{\selectlanguage {en}\bibinfo {title} {Nonlocal stationary probability distributions and escape rates for an active {Ornstein}–{Uhlenbeck} particle}},\ }\href@noop {} {\bibfield  {journal} {\bibinfo  {journal} {J. Stat. Mech. Theor. Exp.}\ }\textbf {\bibinfo {volume} {2020}},\ \bibinfo {pages} {063204} (\bibinfo {year} {2020})}\BibitemShut {NoStop}%
\bibitem [{\citenamefont {Martin}\ \emph {et~al.}(2021)\citenamefont {Martin}, \citenamefont {O'Byrne}, \citenamefont {Cates}, \citenamefont {Fodor}, \citenamefont {Nardini}, \citenamefont {Tailleur},\ and\ \citenamefont {van Wijland}}]{martin_statistical_2021}%
  \BibitemOpen
  \bibfield  {author} {\bibinfo {author} {\bibfnamefont {D.}~\bibnamefont {Martin}}, \bibinfo {author} {\bibfnamefont {J.}~\bibnamefont {O'Byrne}}, \bibinfo {author} {\bibfnamefont {M.~E.}\ \bibnamefont {Cates}}, \bibinfo {author} {\bibfnamefont {E.}~\bibnamefont {Fodor}}, \bibinfo {author} {\bibfnamefont {C.}~\bibnamefont {Nardini}}, \bibinfo {author} {\bibfnamefont {J.}~\bibnamefont {Tailleur}},\ and\ \bibinfo {author} {\bibfnamefont {F.}~\bibnamefont {van Wijland}},\ }\bibfield  {title} {\bibinfo {title} {Statistical mechanics of active {Ornstein}-{Uhlenbeck} particles},\ }\href@noop {} {\bibfield  {journal} {\bibinfo  {journal} {Phys. Rev. E}\ }\textbf {\bibinfo {volume} {103}},\ \bibinfo {pages} {032607} (\bibinfo {year} {2021})}\BibitemShut {NoStop}%
\bibitem [{\citenamefont {Martin}\ and\ \citenamefont {Pirey}(2021)}]{martin_aoup_2021}%
  \BibitemOpen
  \bibfield  {author} {\bibinfo {author} {\bibfnamefont {D.}~\bibnamefont {Martin}}\ and\ \bibinfo {author} {\bibfnamefont {T.~A.~d.}\ \bibnamefont {Pirey}},\ }\bibfield  {title} {{\selectlanguage {en}\bibinfo {title} {{AOUP} in the presence of {Brownian} noise: a perturbative approach}},\ }\href@noop {} {\bibfield  {journal} {\bibinfo  {journal} {J. Stat. Mech. Theor. Exp.}\ }\textbf {\bibinfo {volume} {2021}},\ \bibinfo {pages} {043205} (\bibinfo {year} {2021})}\BibitemShut {NoStop}%
\bibitem [{\citenamefont {Nguyen}\ \emph {et~al.}(2021)\citenamefont {Nguyen}, \citenamefont {Wittmann},\ and\ \citenamefont {Löwen}}]{nguyen_active_2021}%
  \BibitemOpen
  \bibfield  {author} {\bibinfo {author} {\bibfnamefont {G.~H.~P.}\ \bibnamefont {Nguyen}}, \bibinfo {author} {\bibfnamefont {R.}~\bibnamefont {Wittmann}},\ and\ \bibinfo {author} {\bibfnamefont {H.}~\bibnamefont {Löwen}},\ }\bibfield  {title} {{\selectlanguage {en}\bibinfo {title} {Active {Ornstein}–{Uhlenbeck} model for self-propelled particles with inertia}},\ }\href@noop {} {\bibfield  {journal} {\bibinfo  {journal} {J. Phys. Condens. Matter.}\ }\textbf {\bibinfo {volume} {34}},\ \bibinfo {pages} {035101} (\bibinfo {year} {2021})}\BibitemShut {NoStop}%
\bibitem [{\citenamefont {Crisanti}\ and\ \citenamefont {Paoluzzi}(2023)}]{crisanti_most_2023}%
  \BibitemOpen
  \bibfield  {author} {\bibinfo {author} {\bibfnamefont {A.}~\bibnamefont {Crisanti}}\ and\ \bibinfo {author} {\bibfnamefont {M.}~\bibnamefont {Paoluzzi}},\ }\bibfield  {title} {\bibinfo {title} {Most probable path of active {Ornstein}-{Uhlenbeck} particles},\ }\href@noop {} {\bibfield  {journal} {\bibinfo  {journal} {Phys. Rev. E}\ }\textbf {\bibinfo {volume} {107}},\ \bibinfo {pages} {034110} (\bibinfo {year} {2023})}\BibitemShut {NoStop}%
\bibitem [{\citenamefont {Dutta}(2023)}]{dutta_most_2023}%
  \BibitemOpen
  \bibfield  {author} {\bibinfo {author} {\bibfnamefont {S.}~\bibnamefont {Dutta}},\ }\bibfield  {title} {\bibinfo {title} {Most probable paths for active {Ornstein}-{Uhlenbeck} particles},\ }\href@noop {} {\bibfield  {journal} {\bibinfo  {journal} {Phys. Rev. E}\ }\textbf {\bibinfo {volume} {107}},\ \bibinfo {pages} {054130} (\bibinfo {year} {2023})}\BibitemShut {NoStop}%
\bibitem [{\citenamefont {Fritz}\ and\ \citenamefont {Seifert}(2023)}]{fritz_thermodynamically_2023}%
  \BibitemOpen
  \bibfield  {author} {\bibinfo {author} {\bibfnamefont {J.~H.}\ \bibnamefont {Fritz}}\ and\ \bibinfo {author} {\bibfnamefont {U.}~\bibnamefont {Seifert}},\ }\bibfield  {title} {{\selectlanguage {en}\bibinfo {title} {Thermodynamically consistent model of an active {Ornstein}–{Uhlenbeck} particle}},\ }\href@noop {} {\bibfield  {journal} {\bibinfo  {journal} {J. Stat. Mech. Theor. Exp.}\ }\textbf {\bibinfo {volume} {2023}},\ \bibinfo {pages} {093204} (\bibinfo {year} {2023})}\BibitemShut {NoStop}%
\bibitem [{\citenamefont {Hueschen}\ \emph {et~al.}(2024)\citenamefont {Hueschen}, \citenamefont {Segev-Zarko}, \citenamefont {Chen}, \citenamefont {LeGros}, \citenamefont {Larabell}, \citenamefont {Boothroyd}, \citenamefont {Phillips},\ and\ \citenamefont {Dunn}}]{hueschen_emergent_2024}%
  \BibitemOpen
  \bibfield  {author} {\bibinfo {author} {\bibfnamefont {C.~L.}\ \bibnamefont {Hueschen}}, \bibinfo {author} {\bibfnamefont {L.-a.}\ \bibnamefont {Segev-Zarko}}, \bibinfo {author} {\bibfnamefont {J.-H.}\ \bibnamefont {Chen}}, \bibinfo {author} {\bibfnamefont {M.~A.}\ \bibnamefont {LeGros}}, \bibinfo {author} {\bibfnamefont {C.~A.}\ \bibnamefont {Larabell}}, \bibinfo {author} {\bibfnamefont {J.~C.}\ \bibnamefont {Boothroyd}}, \bibinfo {author} {\bibfnamefont {R.}~\bibnamefont {Phillips}},\ and\ \bibinfo {author} {\bibfnamefont {A.~R.}\ \bibnamefont {Dunn}},\ }\bibfield  {title} {{\selectlanguage {en}\bibinfo {title} {Emergent actin flows explain distinct modes of gliding motility}},\ }\href@noop {} {\bibfield  {journal} {\bibinfo  {journal} {Nature Physics}\ ,\ \bibinfo {pages} {1}} (\bibinfo {year} {2024})}\BibitemShut {NoStop}%
\bibitem [{\citenamefont {Bradbury}\ \emph {et~al.}(2018)\citenamefont {Bradbury}, \citenamefont {Frostig}, \citenamefont {Hawkins}, \citenamefont {Johnson}, \citenamefont {Leary}, \citenamefont {Maclaurin}, \citenamefont {Necula}, \citenamefont {Paszke}, \citenamefont {VanderPlas}, \citenamefont {Wanderman-Milne},\ and\ \citenamefont {Zhang}}]{bradbury_jax_2018}%
  \BibitemOpen
  \bibfield  {author} {\bibinfo {author} {\bibfnamefont {J.}~\bibnamefont {Bradbury}}, \bibinfo {author} {\bibfnamefont {R.}~\bibnamefont {Frostig}}, \bibinfo {author} {\bibfnamefont {P.}~\bibnamefont {Hawkins}}, \bibinfo {author} {\bibfnamefont {M.~J.}\ \bibnamefont {Johnson}}, \bibinfo {author} {\bibfnamefont {C.}~\bibnamefont {Leary}}, \bibinfo {author} {\bibfnamefont {D.}~\bibnamefont {Maclaurin}}, \bibinfo {author} {\bibfnamefont {G.}~\bibnamefont {Necula}}, \bibinfo {author} {\bibfnamefont {A.}~\bibnamefont {Paszke}}, \bibinfo {author} {\bibfnamefont {J.}~\bibnamefont {VanderPlas}}, \bibinfo {author} {\bibfnamefont {S.}~\bibnamefont {Wanderman-Milne}},\ and\ \bibinfo {author} {\bibfnamefont {Q.}~\bibnamefont {Zhang}},\ }\href@noop {} {\bibinfo {title} {{JAX}: composable transformations of {Python}+{NumPy} programs}} (\bibinfo {year} {2018})\BibitemShut {NoStop}%
\bibitem [{\citenamefont {Kidger}(2021)}]{kidger_neural_2021}%
  \BibitemOpen
  \bibfield  {author} {\bibinfo {author} {\bibfnamefont {P.}~\bibnamefont {Kidger}},\ }\emph {\bibinfo {title} {On {Neural} {Differential} {Equations}}},\ \href@noop {} {\bibinfo {type} {{PhD} {Thesis}}},\ \bibinfo  {school} {University of Oxford} (\bibinfo {year} {2021})\BibitemShut {NoStop}%
\bibitem [{\citenamefont {Olsen}\ and\ \citenamefont {L{\"o}wen}(2024)}]{olsen2024optimal}%
  \BibitemOpen
  \bibfield  {author} {\bibinfo {author} {\bibfnamefont {K.~S.}\ \bibnamefont {Olsen}}\ and\ \bibinfo {author} {\bibfnamefont {H.}~\bibnamefont {L{\"o}wen}},\ }\bibfield  {title} {\bibinfo {title} {Optimal diffusion of chiral active particles with strategic reorientations},\ }\href@noop {} {\bibfield  {journal} {\bibinfo  {journal} {Phys. Rev. E}\ }\textbf {\bibinfo {volume} {110}},\ \bibinfo {pages} {064606} (\bibinfo {year} {2024})}\BibitemShut {NoStop}%
\bibitem [{\citenamefont {Wan}\ and\ \citenamefont {Goldstein}(2014)}]{wan_rhythmicity_2014}%
  \BibitemOpen
  \bibfield  {author} {\bibinfo {author} {\bibfnamefont {K.~Y.}\ \bibnamefont {Wan}}\ and\ \bibinfo {author} {\bibfnamefont {R.~E.}\ \bibnamefont {Goldstein}},\ }\bibfield  {title} {\bibinfo {title} {Rhythmicity, {Recurrence}, and {Recovery} of {Flagellar} {Beating}},\ }\href@noop {} {\bibfield  {journal} {\bibinfo  {journal} {Phys. Rev. Lett.}\ }\textbf {\bibinfo {volume} {113}},\ \bibinfo {pages} {238103} (\bibinfo {year} {2014})}\BibitemShut {NoStop}%
\bibitem [{\citenamefont {Ghosh}\ and\ \citenamefont {Fischer}(2009)}]{ghosh_controlled_2009}%
  \BibitemOpen
  \bibfield  {author} {\bibinfo {author} {\bibfnamefont {A.}~\bibnamefont {Ghosh}}\ and\ \bibinfo {author} {\bibfnamefont {P.}~\bibnamefont {Fischer}},\ }\bibfield  {title} {\bibinfo {title} {Controlled {Propulsion} of {Artificial} {Magnetic} {Nanostructured} {Propellers}},\ }\href@noop {} {\bibfield  {journal} {\bibinfo  {journal} {Nano Lett.}\ }\textbf {\bibinfo {volume} {9}},\ \bibinfo {pages} {2243} (\bibinfo {year} {2009})}\BibitemShut {NoStop}%
\bibitem [{\citenamefont {Yamamoto}\ and\ \citenamefont {Sano}(2017)}]{yamamoto2017chirality}%
  \BibitemOpen
  \bibfield  {author} {\bibinfo {author} {\bibfnamefont {T.}~\bibnamefont {Yamamoto}}\ and\ \bibinfo {author} {\bibfnamefont {M.}~\bibnamefont {Sano}},\ }\bibfield  {title} {\bibinfo {title} {Chirality-induced helical self-propulsion of cholesteric liquid crystal droplets},\ }\href@noop {} {\bibfield  {journal} {\bibinfo  {journal} {Soft Matter}\ }\textbf {\bibinfo {volume} {13}},\ \bibinfo {pages} {3328} (\bibinfo {year} {2017})}\BibitemShut {NoStop}%
\bibitem [{\citenamefont {Llacer-Wintle}\ \emph {et~al.}(2021)\citenamefont {Llacer-Wintle}, \citenamefont {Rivas-Dapena}, \citenamefont {Chen}, \citenamefont {Pellicer}, \citenamefont {Nelson}, \citenamefont {Puigmart{\'\i}-Luis},\ and\ \citenamefont {Pan{\'e}}}]{llacer2021biodegradable}%
  \BibitemOpen
  \bibfield  {author} {\bibinfo {author} {\bibfnamefont {J.}~\bibnamefont {Llacer-Wintle}}, \bibinfo {author} {\bibfnamefont {A.}~\bibnamefont {Rivas-Dapena}}, \bibinfo {author} {\bibfnamefont {X.-Z.}\ \bibnamefont {Chen}}, \bibinfo {author} {\bibfnamefont {E.}~\bibnamefont {Pellicer}}, \bibinfo {author} {\bibfnamefont {B.~J.}\ \bibnamefont {Nelson}}, \bibinfo {author} {\bibfnamefont {J.}~\bibnamefont {Puigmart{\'\i}-Luis}},\ and\ \bibinfo {author} {\bibfnamefont {S.}~\bibnamefont {Pan{\'e}}},\ }\bibfield  {title} {\bibinfo {title} {Biodegradable small-scale swimmers for biomedical applications},\ }\href@noop {} {\bibfield  {journal} {\bibinfo  {journal} {Adv. Mater.}\ }\textbf {\bibinfo {volume} {33}},\ \bibinfo {pages} {2102049} (\bibinfo {year} {2021})}\BibitemShut {NoStop}%
\bibitem [{\citenamefont {Leung}\ \emph {et~al.}(2014)\citenamefont {Leung}, \citenamefont {Rould}, \citenamefont {Konradt}, \citenamefont {Hunter},\ and\ \citenamefont {Ward}}]{leung_disruption_2014}%
  \BibitemOpen
  \bibfield  {author} {\bibinfo {author} {\bibfnamefont {J.~M.}\ \bibnamefont {Leung}}, \bibinfo {author} {\bibfnamefont {M.~A.}\ \bibnamefont {Rould}}, \bibinfo {author} {\bibfnamefont {C.}~\bibnamefont {Konradt}}, \bibinfo {author} {\bibfnamefont {C.~A.}\ \bibnamefont {Hunter}},\ and\ \bibinfo {author} {\bibfnamefont {G.~E.}\ \bibnamefont {Ward}},\ }\bibfield  {title} {{\selectlanguage {en}\bibinfo {title} {Disruption of {TgPHIL1} {Alters} {Specific} {Parameters} of {Toxoplasma} gondii {Motility} {Measured} in a {Quantitative}, {Three}-{Dimensional} {Live} {Motility} {Assay}}},\ }\href@noop {} {\bibfield  {journal} {\bibinfo  {journal} {PLOS ONE}\ }\textbf {\bibinfo {volume} {9}},\ \bibinfo {pages} {e85763} (\bibinfo {year} {2014})}\BibitemShut {NoStop}%
\end{thebibliography}%

\onecolumngrid
\section*{End Matter}
\twocolumngrid

\setcounter{figure}{0}
\renewcommand{\thefigure}{A\arabic{figure}}

\begin{figure}
    \centering
    \includegraphics[width=0.666\columnwidth]{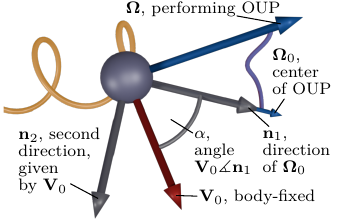}
    \caption{Model schematics. The translational velocity $\Vo$ is fixed in the body frame, but the angular velocity $\W$ performs an Ornstein-Uhlenbeck process (OUP) centered around the body fixed $\WoVec$. The body coordinates are given by the vectors $\nW$, the direction of the average angular velocity $\WoVec$, and $\nV$, chosen such that $\Vo$ is in the plane spanned by $\nW$,$\nV$, with an angle of $\Van$ between $\Vo$ and $\nW$. See also Supp. Movies 1-3.
    }
    \label{fig:Schematics}
\end{figure}

{\it Appendix A - Experimental details:}

Malaria is caused by unicellular parasites of the genus {\it Plasmodium}. During its complex life cycle, the sporozoite, a 10~$\upmu$m long, crescent shaped motile stage of the parasite, is transmitted from a mosquito during its blood meal and utilizes rapid gliding motility to migrate in the skin to enter blood vessels.
The experimental setup \cite{ripp_malaria_2021} consists of a soft and porous polyacrylamide hydrogel serving as 3D substrate mimicking the host skin. A mosquito salivary gland infected by \textit{P. berghei} sporozoites that express a fluorescent protein in their cytoplasm was placed on top of the gel, such that sporozoites could invade into the gel at high numbers. 3D sporozoite migration was observed using spinning disc confocal microscopy, which allowed us to follow the rapid migration of hundreds of parasites.

The microscopy results were processed by an automated image analysis pipeline. Standard filtering and registration approaches were combined with a custom build deconvolution and tracking pipeline to obtain 3D trajectories of individual sporozoites.

Experimentally observed sporozoites do not move at a constant speed, and even their average speed can vary between different parasites by a factor of 3. For this analysis, we resampled the trajectories by fitting a Fourier series (similar to \cite{leung_disruption_2014}) and assuming a constant speed of 1~$\upmu$m/s. \new{We used 140 trajectories with lengths between 100 and 300~$\upmu$m.} The fit of the MSD in \Fig\ref{fig:Fig5} results in the parameters given in Table \ref{tab:fit_results}.

\begin{figure}[b]
    \centering
    \includegraphics[width=\columnwidth]{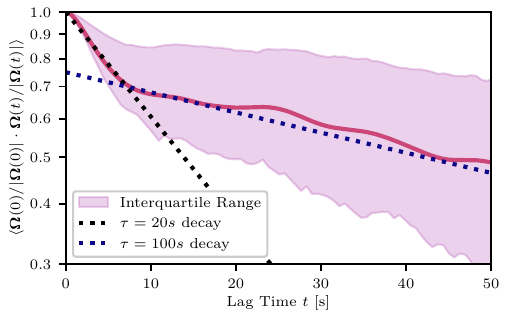}
    \caption{\new{The autocorrelation of the normalized Darboux vector, $\langle\mathbf{\Omega}(0)/|\mathbf{\Omega}(0)|\cdot\mathbf{\Omega}(t)/|\mathbf{\Omega}(t)|\rangle$, as estimated from experimental trajectories, displays a decay of autocorrelation with a fast ($\tau=$~20~s) and a slow ($\tau=$~100~s) regime. The shaded area displays the 25-75\% interquartile range and the solid line is the average.}}
    \label{fig:A2}
\end{figure}

To estimate the correlation timescale of the angular velocity, we can obtain the estimated vector $\W\att$ from the trajectories as the Darboux vector of the Frenet frame, which can be derived from the fitted Fourier series.
While the modulus shows a relatively noisy behaviour, the direction decays on two clearly separated 
time scales as shown in \Fig\ref{fig:A2}. The large time scale of $\tau\approx$~100~s describes the decorrelation of the helical axis. Additionally, a second, shorter timescale is visible, as expected from the Ornstein-Uhlenbeck process (Gaussian white noise would lead to decay with only a single timescale). This short time scale of $\tau=$~20~s represents the time scale on which the axis of rotation of the internal force generating apparatus fluctuates during motion. It is notably smaller than the time scale $1/\OUth$ as extracted from the MSD fit, which would give the time scale of decay for the full $\W$ in the OUP if $\nW$ would be fixed.
Note, however, that the direction of angular velocity $\W/|\W|$ relative to the moving center of the OUP follows a more complicated decay law. Additionally, it is likely that for the malaria parasite the assumed isotropy of the OUP is not exactly true, and the magnitude of the angular velocity fluctuates faster then the direction.

\begin{table}[t]
    \centering
    \caption{Results for fitting the model to the MSD computed from experimentally observed and resampled trajectories of malaria parasites.}
    \begin{tabularx}{\columnwidth}{>{\raggedright\arraybackslash}l>{\centering\arraybackslash}X>{\centering\arraybackslash}X>{\centering\arraybackslash}X>{\centering\arraybackslash}X}
        \toprule
        Parameter & $\Omega_0$ & $\alpha$ & $\OUth$ & $\OUamp$  \\ 
        \midrule
        Fit Result & 0.285~1/s & 0.926 & 1.078~1/s & 0.154~1/s$^{3/2}$  \\ 
        \bottomrule
    \end{tabularx}
    \label{tab:fit_results}
\end{table}

{\it Appendix B - Moment Closure of the Rotational Problem:}

The dynamic equations for the first four moments expanding the rotational problem (\Eq\ref{eq:dOm}-\ref{eq:dN1}) are
\begin{align}
    \D\expv{\W} =&\ -\OUth \left(\expv{\W}-\Wo\expv{\nW}\right)\Dt \label{eq:hier1full}\\
    \D\expv{\nW} =&\ \expv{\W\times\nW}\Dt\label{eq:hier2full}\\
    \D\expv{\W\times\nW} =&\ -\OUth \expv{\W\times\nW}\Dt
    + \expv{\W\times(\W\times\nW)}\Dt\label{eq:hier3full}
\end{align}
and
\begin{align*}
\D\expv{\W\times(\W\times\nW)} =&\ - 2 \OUth \expv{\W\times(\W\times\nW)} \Dt\nonumber\\
    &\ + \expv{\W\times\left(\W\times(\W\times\nW)\right)}\Dt\nonumber\\
    &\ + \OUth\Wo\expv{\nW\times(\W\times\nW)}\Dt \nonumber\\
    &\ + \OUamp^2\expv{\dnoise\times(\dnoise\times\nW)}\ .
\end{align*}
To obtain an analytical solution, we apply moment closure by
approximating the second and third term in this equation. The truncation cannot be performed 
earlier due to the necessity of retaining terms up to second order in $\Omega$ to properly account for the 
effect of noise.
First, assuming $\expv{\W^2 A}\approx\Wo^2\expv{A}$, 
we get for the second term
$\expv{\W\times\left(\W\times(\W\times\nW)\right)} \approx -\Wo^2\expv{\W\times\nW}$.
For the third term, 
upon replacing $\W$ with $\W^\perp$, its component
perpendicular to $\nW$, the relevant 
contribution for the cross product, and 
applying similar logic as before but 
with $|\nW|=1$, we get
    $\expv{\nW\times(\W\times\nW)} 
    = \expv{\W^\perp}= \expv{\W-(\W\cdot\nW)\nW} \approx \expv{\W} - \Wo\expv{\nW}$.
For the last approximation, we assumed 
that the variance of the OU-process is
small, such that $\W$ stays close 
to its average $\Wo\nW$.
The noise term can be explicitly computed, and we can close the hierarchy by rewriting its fourth equation as
\begin{align}\label{eq:hier4trunc} 
    &\D\expv{\W\times(\W\times\nW)} = - 2 \OUth \expv{\W\times(\W\times\nW)} \Dt\\
    &\  - \Wo^2 \expv{\W\times\nW}\Dt  + \OUth\Wo\left(\expv{\W} - \Wo\expv{\nW}\right)\Dt - 2 \OUamp^2\expv{\nW}\Dt .\nonumber
\end{align}

The truncation presented above breaks down for $\Wo=0$, 
because the equation for $\W$ decouples 
from the rest, and $\expv{\W}$ is 
dominated by the mean squared displacement (MSD) of the OUP
instead of $\Wo$. However, in the limit of small noise ($\OUamp^2\ll\OUth^3$), the previously derived eigenvalue (\Eq\ref{eq:eigval}) has a well-defined limit
\begin{equation}
    \lambda\Big|_{\Wo=0}\approx -\frac{\OUamp^2}{\OUth^2} \ .
\end{equation}
Following through the previous derivation of \Eq\ref{eq:hier4trunc} for the case of $\Wo=0$, and truncating $\expv{\W\times\left(\W\times(\W\times\nW)\right)} \approx -\OUamp^2/(2\OUth)\expv{\W\times\nW}$, using the MSD of the OUP instead of $\Wo^2$, we find the same result. Therefore, even if the original derivation is not valid, the eigenvalue as written correctly includes the $\Wo\to0$ limiting case. This is also confirmed by the numerical simulations for the $\Wo=0$ case in \Fig\ref{fig:Trans}.\\

{\it Appendix C - Derivation of MSD:} 

The MSD can now be obtained by first computing it from the formal solution
\begin{equation}
    \pos\att = \int_0^t\Dt\left(\cos{(\Van)}\nW+\sin{(\Van)}\nV)\right) \ ,
\end{equation}
which yields
\begin{equation}
\begin{split}
    \expv{\pos\att^2} = \int_0^t\D s_1 &\int_0^t \D s_2 \Big(\cos^2{(\Van)}\expv{\nW(s_1)\cdot\nW(s_2)}\\
    &+\sin^2{(\Van)}\expv{\nV(s_1)\cdot\nV(s_2)}\Big)\ ,
\end{split}
\end{equation}
where the mixed terms vanish, as due to their perpendicularity and the rotational symmetry the expectation value of their scalar products has to be zero even if evaluated at different times. The remaining correlation functions can be directly obtained from the solutions obtained for $\expv{\nW}$ and $\expv{\nV}$ as
\begin{align}
    \expv{\nW(s_1)\cdot\nW(s_2)} =&\ e^{\lambda|s_1-s_2|}\ ,\\
    \expv{\nV(s_1)\cdot\nV(s_2)} =&\ \cos{(\Wo |s_1-s_2|)} e^{2\lambda|s_1-s_2|}\ ,
\end{align}
such that we finally compute the MSD by simple integration.\\

\end{document}


\setcounter{equation}{0}
\renewcommand{\theequation}{S\arabic{equation}}

\newcommand{\red}[1]{{\leavevmode\color{black}#1}}
\newcommand{\new}[1]{{\leavevmode\color{black}#1}}

\newcommand{\Eq}{Eq.~}
\newcommand{\Fig}{Fig.~}

\newcommand{\W}{\mathbf{\Omega}}
\newcommand{\nW}{\mathbf{n}_1}
\newcommand{\nV}{\mathbf{n}_2}
\newcommand{\nT}{\mathbf{n}_3}
\newcommand{\nF}{\mathbf{n}_4}
\newcommand{\pos}{\mathbf{r}}
\newcommand{\att}{_{(t)}}
\newcommand{\ats}{_{(s)}}
\newcommand{\at}[1]{_{(#1)}}
\newcommand{\expv}[1]{\left\langle #1 \right\rangle}
\newcommand{\D}{\text{d}}
\newcommand{\Dt}{\text{d}t}

\newcommand{\Vo}{\mathbf{V}_0}
\newcommand{\Vomag}{V_0}
\newcommand{\WoVec}{\mathbf{\Omega}_0}
\newcommand{\Wo}{\Omega_0}
\newcommand{\wvec}{\mathbf{\omega}}
\newcommand{\wmag}{\omega}
\newcommand{\norm}[1]{\left\lVert#1\right\rVert}

\newcommand{\rot}{\mathcal{R}}
\newcommand{\dnoise}{\text{d}\mathbf{\Lambda}}
\newcommand{\dscnoise}{\text{d}\Lambda}
\newcommand{\inbody}{^{\text{body}}}
\newcommand{\inlab}{^{\text{lab}}}

\newcommand{\OUth}{k}
\newcommand{\OUamp}{h}

\newcommand{\Van}{\alpha}


\title{Supplemental Material:\\
Three-dimensional chiral active Ornstein-Uhlenbeck model\\
for helical motion of microorganisms}

\author{Leon Lettermann}
\author{Falko Ziebert}
\affiliation{Institute for Theoretical Physics, Heidelberg University, Philosophenweg 19, 69120 Heidelberg, Germany}
\affiliation{Bioquant-Center, Heidelberg University, Im Neuenheimer Feld 267, 69120 Heidelberg, Germany}
\author{Mirko Singer}
\affiliation{Parasitology, Center for Infectious Diseases, Heidelberg University, Im Neuenheimer Feld 344, 69120 Heidelberg}
\author{Friedrich Frischknecht}
\affiliation{Parasitology, Center for Infectious Diseases, Heidelberg University, Im Neuenheimer Feld 344, 69120 Heidelberg}
\affiliation{German Center for Infection Research (DZIF), Partner Site Heidelberg, 69120 Heidelberg, Germany}
\author{Ulrich S. Schwarz}
\affiliation{Institute for Theoretical Physics, Heidelberg University, Philosophenweg 19, 69120 Heidelberg, Germany}
\affiliation{Bioquant-Center, Heidelberg University, Im Neuenheimer Feld 267, 69120 Heidelberg, Germany}

\date{\today}

\maketitle

\onecolumngrid
\section*{Supplementary figures}
\twocolumngrid

\renewcommand\thefigure{S\arabic{figure}} 

\begin{figure}[h]
    \centering
    \includegraphics[width=\linewidth]{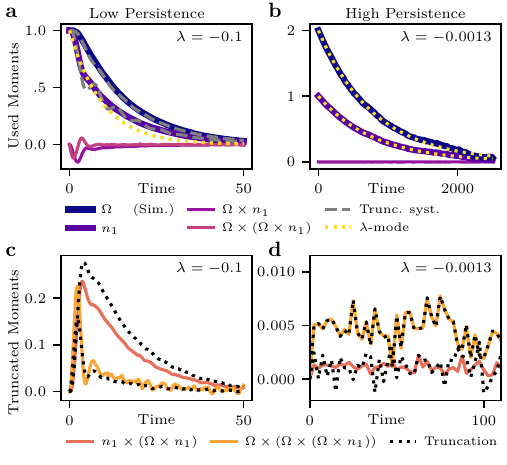}
    \caption{\textbf{a}: Time course of different moments obtained 
    by averaging numerical simulation (\Eq2-5) in comparison with numerical solution of the truncated system (\Eq12-15, gray dashed lines)
    and analytical approximation predicting exponential
    decay with eigenvalue $\lambda$ (\Eq6, yellow dotted lines).
    Parameter values: potential strength $\OUth=$~0.2, noise amplitude 
    $\OUamp=$~0.3, angular speed $\Wo=$~1, angle $\Van=\pi/6$.
    \textbf{b}: Same as a, but now for $\OUth=$~2, $\OUamp=$~0.1 and $\Wo=$~2, i.e.~much reduced noise and faster turning. 
    Here, the agreement between simulations and theory is even better.
    \textbf{c}: The truncated moments and the approximation used for truncation in \Eq15.
    \textbf{d}: Same as c at parameters from b. Here, the truncated moments are already very small, and even at 20.000 simulations the averages have not yet completely converged. Nonetheless, it is apparent that the approximations used for truncation work well.
    }
    \label{fig:TruncatedMoments}
\end{figure}

\begin{figure}[h]
    \centering
    \includegraphics[width=\linewidth]{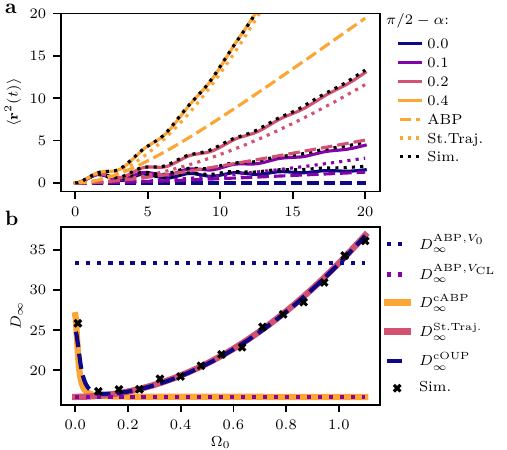}
    \caption{\textbf{a}: Mean squared displacement for $\Van$ close to $\pi/2$, such that the particles are close to describing circles, with $\OUth=$~1, $\OUamp=$~0.5, $\Wo=$~2. Black dotted lines are averages from numerical simulations. Dashed colored lines are the MSD as estimated for the active Brownian particle \red{(ABP) with rotational diffusion derived from the variance of the Ornstein-Uhlenbeck process \new{(\Eq\ref{eq:MSD_ABP}, \cite{howse_self-motile_2007})}. Dotted lines are the MSD for the helical centerline with the persistence derived from stochastic trajectory models \new{(\Eq\ref{eq:DOmStochTraj}, \cite{friedrich_steering_2009})}.}
    \red{\textbf{b}: Long-time effective diffusion $D_\infty$ for different models described in Supplemental Note 1 \new{(ABP, \Eq\ref{eq:DInfABP}; chiral ABP, \Eq\ref{eq:cABPDinf}, \cite{kirkegaard2016motility}; stochastic trajectories, \Eq\ref{eq:DInfStochTraj})}, compared to the result found in \Eq10 for the chiral OUP. Parameter values: potential strength $\OUth=$~1, noise amplitude 
    $\OUamp=$~0.1, angular speed $\Wo=$~1, angle $\Van=\pi/4$.}
    }
    \label{fig:MSDwoSin}
\end{figure}

\clearpage

\onecolumngrid
\section*{Supplemental Note 1: Comparison with related models}
\twocolumngrid

In this supplementary note, we compare the results of our model with related results from the literature.\par
\textbf{Active Brownian particles} (ABPs) are the most common active particle model, and are usually written down with a translational and rotational noise, the latter given by a rotational diffusion $D_\Omega$ \cite{howse_self-motile_2007, sevilla2016diffusion}. Disregarding the translational noise (to compare to our approach), the MSD in 3D reads
\begin{equation}\label{eq:MSD_ABP}
    \expv{\pos^2\att} =  \frac{\Vomag^2}{D_\Omega}t - \frac{\Vomag^2}{2 D_\Omega^2}\left(1-e^{-2D_\Omega t}\right) \,
\end{equation}
which reproduces the $\Van=0$ limit from \Eq13 with $\lambda=-2D_\Omega$.

\textbf{Worm-like chain} (WLC) models \cite{kratky1949rontgenuntersuchung, doi1988theory} describe the configuration of a fixed-length polymer with a given persistence length. Replacing the total length with the traveled distance (using a velocity $\Vomag$), and writing a persistence time $\tau = P/\Vomag$ instead of the persistence length $P$ we recover the MSD of the ABP above, where $\tau = -1/\lambda = 2/D_\Omega$.

The ABP or WLC model can describe the MSD of our theory for $\Van=0$, but deviate for the true helical case ($\Van>0$). This is apparent for the long time limit, where \Eq\ref{eq:MSD_ABP} yields (scaling $\Vomag$ with $\cos\alpha$) $D_\infty^\text{ABP} = \cos^2(\alpha) D_\Omega$, i.e. these theories do not include the translation generated by the imperfect circular motion in the plane perpendicular to the helix, which is relevant for larger $\Van$ as seen from \Fig\ref{fig:MSDwoSin}a.

\textbf{Chiral active Brownian particles} as discussed in \cite{kirkegaard2016motility} include this effect. Rewriting the long-time diffusion from their work by replacing the velocities $v_p=\cos\alpha$ and $v_\omega=\sin\alpha$ it reads (Eq. 4 in \cite{kirkegaard2016motility})
\begin{equation}\label{eq:cABPDinf}
    D_\infty^\text{cABP} = \frac{\cos^2(\alpha)}{4 D_\Omega} + \frac{\sin^2(\alpha) D_\Omega}{4}\left({\textstyle\frac{1}{9D^2_\Omega+\Omega_0^2}+\frac{2}{4D^2_\Omega+\Omega_0^2}}\right),
\end{equation}
which has a similar form as our \Eq10, \red{but considers a 2D projection of the 3D problem because their experiments imaged 2D projections}. However, it decreases monotonically with increasing $\Omega_0$. Hence, importantly, the stabilizing influence of rotation as observed in our model is not present, as expected because their rotational noise is classical rotational diffusion (external, white noise); thus, the rotation cannot act to integrate out part of the noise, \new{which requires the combination of noise being time correlated and internal}.

\red{
\textbf{Arbitrarily shaped active Brownian particles} were studied in 3D in \cite{wittkowski_self-propelled_2012} by starting from the full 6x6 diffusion tensor and subsequently analyzing its symmetry properties dependent on the shape's symmetry. For an orthothropic particle (i.e. possessing three pairwise orthogonal planes of symmetry) and in the absence of noise they analytically solved the resulting helical trajectory as function of active force, torque and shape-dependent drag of the particle.

By numerical simulations including noise they found the expectation value of many stochastic helical trajectories as an exponentially damped helix, or a concho-spiral, as found here analytically in \Eq11. They fitted two exponential decay scales, $\gamma_1=$~0.04 and $\gamma_2=$~0.06 describing the decay of radius and decay in axial direction respectively. Interestingly, identifying these for the Ornstein-Uhlenbeck particle presented here, we find $\gamma_1=-2\lambda$ and $\gamma_2=-\lambda$, and in particular $\gamma_1>\gamma_2$, meaning faster radial than axial decay. This might be caused by the difference of Brownian to Ornstein-Uhlenbeck particle, or by the particular diffusion tensor used to obtain this result in \cite{wittkowski_self-propelled_2012}.

In a related manner to the shape dependence one could study direction dependent internal noise in the OUP, by exchanging $\OUamp$ and $\OUth$ with 3x3 tensors, describing different noise strength in different directions in the body-frame and necessarily being appropriately rotated to the lab frame. This would allow for example different behavior in the magnitude and orientation of the angular velocity, as discussed for the experimental data in Appendix A. Because this would introduce multiplicative noise, similarly to \cite{wittkowski_self-propelled_2012} an analytical solution would be challenging and likely only possible in certain special cases.}

\textbf{Stochastic trajectory models} have been proposed in \cite{friedrich_stochastic_2008, friedrich_steering_2009} to describe helical swimming and chemotaxis of sperm cells. There, the curvature $\kappa\att$ and torsion $\tau\att$ are the main properties of interest, and modeled including stochastic influences of arbitrary, possibly correlated noise. Although in these references Gaussian white noise is used in their explicit solutions, it is possible to approximately solve our OU model in their framework. Rewriting our model with (in the body-frame) fixed, unit-length velocity $\Vo$ and dynamic angular velocity $\W\att=\Wo\nW+\wvec\att$, with $\wvec$ a 3D Ornstein-Uhlenbeck process centered around 0, curvature and torsion at time $t$ are given as
\begin{align}
    \kappa\att =&\ \norm{\W\att\times\Vo}=\norm{\Wo\sin\Van\nT+\wvec\att\times\Vo}\\
    \tau\att =&\ \W\att\cdot\Vo = \Wo\sin\Van+\wvec\att\cdot\Vo
\end{align}
We can then decompose the 3D OUP $\wvec\att$ into two independent 1D processes, $\wmag\att^\parallel$ parallel to $\Vo$, and $\wmag\att^\perp$ perpendicular to both $\Vo$ and $\nT$. The latter is one of two directions relevant for the curvature, but the leading-order contribution. Expanding the norm we find
\begin{align}
    \kappa\att = \Wo\sin\Van + \wmag^\perp\att\ , &&\tau\att = \Wo\cos\Van + \wmag^\parallel\att\ .
\end{align}
Hence, in leading order we find both the curvature and the torsion being subject to independent Ornstein-Uhlenbeck noises. Using the power spectrum of the OUP \cite{gardiner1985handbook}, we can evaluate the combined power spectrum of curvature and torsion noise ($\tilde{S}_2(\wmag)$ in \cite{friedrich_steering_2009}), as its value at $\Wo$ is proportional to the rotational diffusion of the helical centerline (\Eq(4) in \cite{friedrich_steering_2009}):
\begin{equation}\label{eq:PowerSpec}
    \tilde{S}_2(\Wo) = \frac{h^2}{k^2+\Wo^2}
\end{equation}
Therefore, we recover the power spectrum of an Ornstein-Uhlenbeck process. Furthermore, this result is (up to sign) identical to a low noise expansion $h\ll1$ of the eigenvalue $\lambda$ found in \Eq6:
\red{
\begin{align*}
    \lim_{h\to0}\lambda =&\ \lim_{h\to0}-\frac{\OUamp^2+\Wo^2 \OUth+\OUth^3-\sqrt{\OUth^2 \left(\Wo^2+\OUth^2\right)^2-\OUamp^4}}{\Wo^2+\OUth^2}\\
    =&\ -\frac{\OUamp^2+\Wo^2 \OUth+\OUth^3-\OUth \left(\Wo^2+\OUth^2\right)-\mathcal{O}\left(\OUamp^4\right)}{\Wo^2+\OUth^2}\\
    =&\ -\frac{\OUamp^2}{\Wo^2+\OUth^2}+\mathcal{O}\left(\OUamp^4\right)
\end{align*}}

Thus, we can (to leading order) recover our results for the orientation decorrelation (or equivalently the persistence of the helix center line) from the moment closure procedure by solving our model with the approach from \cite{friedrich_steering_2009}. \new{Note that this only describes the long-term behaviour of the center line, and does not include the oscillations introduced by helical turning relevant at shorter times (cf. \Fig\ref{fig:MSDwoSin}) and necessary to describe the experimental data.}
\new{Furthermore, we can use $\tilde{S}_2(\Wo)$, which encodes the variance of the OUP at the frequency of rotation, to rewrite the eigenvalue $\lambda$ as
\begin{align*}
    \lambda = - \left(\OUth + \tilde{S}_2(\Wo) - \sqrt{\OUth^2-\tilde{S}_2(\Wo)^2}\right),
\end{align*}
illustrating how the decay described by $\lambda$ is a combination of the noise itself (described by $\OUth$) and its interaction with the rotation (described by $\tilde{S}_2(\Wo)$).
}

\red{
\textbf{Limit of OUP as Gaussian White Noise and direct comparison.}
For a direct comparison of the long time diffusion dependency on the angular speed $\Wo$ between the different models, we first look at the Brownian-like limit of the OUP, that is taking $\OUth\to\infty$ and $\OUamp\to\infty$ simultaneously, while keeping the variance $\sigma^2 = \tfrac{h^2}{2k}$ constant. This leads to vanishing correlation time $1/\OUth$ of the Ornstein-Uhlenbeck process, and subsequently $\W$ as a random variable becomes Gaussian white noise, around its center $\Wo\nW$ with fixed variance and delta peak time correlation (see \cite{hanggi1994colored}),
\begin{align*}
    &\quad\W\att\sim \mathcal{N}\left(\Wo\nW(t), \sigma^2\right)\\
    &\left\langle\left(\W\att-\Wo\nW(t)\right)_i \left(\W\ats-\Wo\nW(s)\right)_j\right\rangle = \delta_{ij}\delta_{(t-s)}\ .
\end{align*}
This means $\W$ becomes a non-continuous process, which \new{(ignoring drift)} can be understood as a derivative of Brownian motion as its rescaled integral 
\begin{equation*}
    \mathbf{\Theta}\att = \frac{1}{\sqrt{\OUth}} \int_0^{\OUth t} \W\att = \sigma_*\mathcal{B}_t, \quad \sigma_*^2=2\sigma^2
\end{equation*}
recovers Brownian motion $\mathcal{B}_t$ \cite{bhattacharya_functional_1982}, where the rescaling is necessary, as otherwise the variance of the integral vanishes for $\OUth\to\infty$ .
The resulting system is hence different from a Brownian particle where the angle/orientation performs Brownian motion, but using the rescaling we can identify it with Brownian motion of the orientation with variance $2\sigma^2t/\OUth$, \new{where the additional factor k is arising from matching units of $\mathbf{\Theta}\att$ to represent an angle}.

Utilizing this identification, we can derive the effective long time diffusion $D_\infty$ equivalent to \Eq10 for the different models discussed above. For the classic ABP, we obtain
\begin{equation}\label{eq:DInfABP}
    D_\infty^{\text{ABP}}=\frac{V_0^2}{6 D_\Omega}\quad\text{with}\quad D_\Omega = \frac{\sigma^2}{\new{\OUth}} = \frac{\OUamp^2}{2\OUth^\new{2}}\ ,
\end{equation}
where $V_0$ is either $1$, the particle velocity, or $V_\text{CL}=\cos\Van$, the effective velocity along the centerline.

For the chiral ABP from \cite{kirkegaard2016motility} we take $D_\infty^\text{cABP}$ given in \Eq\ref{eq:cABPDinf}, rescaled by $\tfrac{4}{6}$ to get effective 3D instead of 2D diffusion and also utilizing the same estimated $D_\Omega$ as above.

Lastly, for the stochastic trajectory model from \cite{friedrich_steering_2009} with the decorrelation time for the OUP obtained from the power spectrum in \Eq\ref{eq:PowerSpec}, we take the ABP again but estimate $D_\Omega$ based on the persistence time obtained from the power spectrum as
\begin{align}
    &D_\infty^{\text{St.Traj.}}=\frac{V_0^2}{6 D_\Omega}\label{eq:DInfStochTraj}\\
    &\text{with}\quad D_\Omega = 2\frac{\tilde{S}_2(\Wo)}{4} = \frac{h^2}{2(k^2+\Wo^2)}\ \label{eq:DOmStochTraj}.
\end{align}

\Fig\ref{fig:MSDwoSin}b displays these different results compared to $D_\infty$ for the chiral OUP (\Eq10). The solution for the chiral OUP has a minimum at low (but non-zero) angular speed, which agrees with the estimated behavior of the ABP based on the effective velocity of the helix center line. If the angular speed is approaching zero, both the chiral OUP and the chiral ABP capture that the effective long time diffusion increases: the chirality is hindering the diffusivity, and reducing it makes both solutions approach the ABP solution with the particle velocity $V_0$. Towards higher angular speeds, the chiral OUP as well as the stochastic trajectory model for the OUP capture the increase in persistence due to the rotation, and agree well for these parameters. The stochastic trajectory model describes only the center line, and hence misses the increase of persistence with low angular speed, as this description breaks down. Additionally, if the helix is at low pitch, i.e. $\Van$ close to $\pi/2$, inaccurate circling contributes to the MSD in addition to the motion of the helix centerline, such that the stochastic trajectory model underestimates the MSD for those cases (Dotted lines in \Fig\ref{fig:MSDwoSin}a), while still more accurate than the straight forward ABP estimate.}

\clearpage
\red{
\onecolumngrid
\section*{Supplemental Note 2: Decorrelation of body-frame}
\twocolumngrid
}
We want to find the decorrelation of $\expv{\nV}$ in the regime where $\WoVec$ dominates the noise of the OUP. 
For that, we can use that the rotational problem possesses axial symmetry around the $z$ axis, and that the translational problem is deterministic given a solution of the rotational problem.
Therefore, $z$ components of moments must be invariant under rotating $\nW$ and $\nV$, forcing many expectation values to vanish.
From the previous solution, \Eq13, we obtain
\begin{equation}
    \D\expv{\nW} = \expv{\W\times\nW}\Dt= - \lambda\expv{\nW}\Dt\ .
\end{equation}
The analogous equation for $\nV$ can be expanded by introducing the vector $\nT$, which expands $\nW$ and $\nV$ to an orthonormal basis, and using the Jacobi identity,
\begin{align}
    \D\expv{\nV}/\Dt =&\ \expv{\W\times\nV}= \expv{\W\times\left(\nT\times\nW\right)}\\
    =&\ -\expv{\nT\times\left(\nW\times\W\right)}-\expv{\nW\times\left(\W\times\nT\right)}\nonumber
\end{align}

In the first term, we can write $\nW\times\W=\expv{\nW\times\W}+\mathbf{\Delta}$, where by rotational symmetry $\mathbf{\Delta}$ is isotropic in the $x,y$-plane, and hence $\expv{\nT\times\mathbf{\Delta}}=0$ (at least to first order in $\Delta$), such that with the previous result
\begin{equation}
    -\expv{\nT\times\left(\nW\times\W\right)} = -\expv{\nT\times\left(\lambda\expv{\nW}\right)} = -\lambda\expv{\nV}\ ,
\end{equation}
where the last steps follows a similar argument (introducing a isotropic $\mathbf{\Delta}$ computing the cross product $\nT\times\nW=\nV$ through the averages).
For the second term, similar logic can be applied, yielding
\begin{align}
    -&\expv{\nW\times\left(\W\times\nT\right)}=\ -\expv{\nW\times\left(\W\times\left(-\nV\times\nW\right)\right)}\label{eq:ExpansionJacobi}  \\
    =&\ -\expv{\nW\times\left(\nV\times\left(\nW\times\W\right)\right)} - \expv{\nW\times\left(\nW\times\left(\W\times\nV\right)\right)}\nonumber\\
    =&\ -\lambda\expv{\nV} - \expv{\nW\times\left(\nW\times\left(\W\times\nV\right)\right)}\ .\label{eq:truncN2}
\end{align}
Now, with an analogous argument as used in the truncation in deriving \Eq15 for the last term in \Eq\ref{eq:truncN2} only the $\nW$ orthogonal part of $\Omega\times\nV$ is relevant, which is generated by the $\nW$ parallel component of $\W$. We approximate this as $\W^{\parallel}\approx\Wo\nW$ to obtain
\begin{equation}
    -\expv{\nW\times\left(\nW\times\left(\W\times\nV\right)\right)} = \Wo \expv{\nT}\ ,
\end{equation}
which ultimately combines to
\begin{equation}
    \D\expv{\nV}/\Dt = -2 \lambda \expv{\nV} + \Wo\expv{\nT}\ .
\end{equation}
An analogous equation can be derived for $\expv{\nT}$, giving precisely decay with two times the original eigenvalue $\lambda$ while rotating with angular speed $\Wo$. The two terms in the expansion of \Eq\ref{eq:ExpansionJacobi} can be interpreted as the aforementioned decorrelation of the plane on the one hand, and rotation and decorrelating within the plane on the other hand.

\onecolumngrid
\vspace{1cm}
\section*{Supplementary Movie Captions}
\twocolumngrid

\textbf{Movie 1:}
Animation of simulated trajectories from \Fig2c in the main manuscript. The small vectors denote the body fixed frame (blue: $\nW$, red: $\nV$), while the longer orange vector is the body fixed velocity $\Vo$, and the blue moving vector the angular velocity $\Wo$ performing and Ornstein-Uhlenbeck process around an average aligned with $\nW$. Note that here different from the trajectories in the \Fig2c identical intial orientations were used.

\textbf{Movie 2:}
Animation of simulated trajectories from \Fig2d in the main manuscript. The small vectors denote the body fixed frame (blue: $\nW$, red: $\nV$), while the longer orange vector is the body fixed velocity $\Vo$, and the blue moving vector the angular velocity $\Wo$ performing and Ornstein-Uhlenbeck process around an average aligned with $\nW$. Note that here different from the trajectories in the \Fig2d identical initial orientations were used.

\textbf{Movie 3:}
Animation of simulated trajectories from \Fig5c in the main manuscript. The small vectors denote the body fixed frame (blue: $\nW$, red: $\nV$), while the longer orange vector is the body fixed velocity $\Vo$, and the blue moving vector the angular velocity $\Wo$ performing and Ornstein-Uhlenbeck process around an average aligned with $\nW$. Note that here different from the trajectories in the \Fig5c identical initial orientations were used.

\onecolumngrid
\section*{References}
\twocolumngrid

\bibliography{references}